\documentclass[12pt,a4paper,oneside]{extarticle}
\usepackage{enumitem}
\usepackage{cmap}	
\usepackage[T2A]{fontenc}
\usepackage[utf8]{inputenc}
\usepackage[english]{babel}
\usepackage{amsfonts}
\usepackage{amstext}
\usepackage{amsmath}
\usepackage{amssymb}
\usepackage{indentfirst} 
\usepackage{captcont}

\usepackage[numbers,square,sort]{natbib} 
\bibpunct{(}{)}{,}{a}{,}{,} 

\usepackage{graphics}
\usepackage[pdftex]{graphicx}
\usepackage{color}
\usepackage{lscape} 
\usepackage{bm} 
\usepackage{tikz}
\usepackage{titlesec} 
\usepackage{float} 
\usepackage{longtable}
\usepackage{booktabs}
\usepackage[center]{subfigure}
\usepackage{captcont}
\usepackage{tabularx}
\usepackage{diagbox} 
\usepackage{rotating}
\usepackage{geometry}
\usepackage[super]{nth}
\usepackage{comment} 
\usepackage{textcomp}
\usepackage{graphicx}


\renewcommand\appendix{\par
  \setcounter{section}{0}
  \setcounter{subsection}{0}
  \setcounter{figure}{0}
  \setcounter{table}{0}
    \setcounter{equation}{0}
  \renewcommand\thesection{\Alph{section}}
  \renewcommand\thefigure{\Alph{section}.\arabic{figure}}
  \renewcommand\thetable{А.\arabic{table}}
  \renewcommand\theequation{\arabic{equation}}

}

\definecolor{myred1}{RGB}{255, 0, 0}
\usepackage{tikz}

\begin{document}

\title{Robust Inference on Income Inequality: \\ \textit{t-}Statistic Based Approaches
\footnote{R. Ibragimov and A. Skrobotov' research for this paper was supported by Russian Science Foundation, Project No. 20-78-10113. We thank Mario  Holzner, Clara Mart{\' i}nez-Toledano, Ulrich K. M{\" u}ller, Artem Prokhorov, Alla Salmina, Yulong Wang and the participants at the 
GDN SEE and CIS Research Competition workshops, the Centre for Business Analysis (CEBA, St. Petersburg State University) seminar series, the 14th Annual Global Development Conference of the Global Development Network, the 14th International Conference on Computational and Financial Econometrics and the VIII International Conference on Modern Econometric Tools and Applications for helpful comments and discussions. }}

\author{Rustam Ibragimov$^{a,d}$, Paul Kattuman$^b$, Anton Skrobotov$^{c,d}$ \\
{\small {$^{a}$ Imperial College Business School, Imperial College London}}\\
{\small {$^{b}$ Judge Business School, the University of Cambridge}}\\
{\small {$^{c}$ Russian Presidential Academy of National Economy and Public Administration}}\\
{\small {$^{d}$ Saint Petersburg University}}}
\date{October 29, 2021}
\maketitle
\begin{abstract}
Empirical analyses on income and wealth inequality and those in other fields in economics and finance often face the difficulty that the data is heterogeneous, heavy-tailed or correlated in some unknown fashion. The paper focuses on applications of the recently developed \textit{t}-statistic based robust inference approaches in the analysis of inequality measures and their comparisons under the above problems. Following the approaches, in particular, a robust large sample test on equality of two parameters of interest (e.g., a test of equality of inequality measures in two regions or countries considered) is conducted as follows: The data in the two samples dealt with is partitioned into fixed numbers $q_1, q_2\ge 2$ (e.g., $q_1=q_2=2, 4, 8$) of groups, the parameters (inequality measures dealt with) are estimated for each group, and inference is based on a standard two-sample $t-$test with the resulting $q_1, q_2$ group estimators. Robust $t-$statistic approaches result in valid inference under general conditions that group estimators of parameters (e.g., inequality measures) considered are asymptotically independent, unbiased and Gaussian of possibly different variances, or weakly converge, at an arbitrary rate, to independent scale mixtures of normal random variables. These conditions are typically satisfied in empirical applications even under pronounced heavy-tailedness and heterogeneity and possible dependence in observations. 
The methods dealt with in the paper complement and compare favorably with other inference approaches available in the literature. The use of robust inference approaches is illustrated by an empirical analysis of income inequality measures and their comparisons across different regions in Russia. 


\par \indent \emph{Keywords:} Income inequality, inequality measures, robust inference, heavy-tailedness, Russian economy.
\par \indent \emph{JEL Codes}: C14, D31, D63
\end{abstract}



\section{Introduction} \label{intro}



Empirical analyses on income and wealth inequality and those in other fields in economics and finance often face the difficulty that the data is heterogeneous, heavy-tailed or correlated in some unknown fashion (see, among others, the discussion and reviews in 

Importantly, many studies going back to V. Pareto indicate that income and wealth distributions are heavy-tailed and follow power laws 
\begin{eqnarray} \label{power1} P(X>x)\sim Cx^{-\zeta}, \;\; C>0,\end{eqnarray}
with the tail index $\zeta>0$ (see, among others, the discussion and reviews in \citeauthor{PS}, \citeyear{PS}, \citeauthor{Atkinson}, \citeyear{Atkinson}, \citeauthor{Gabaix1}, \citeyear{Gabaix1}, \citeauthor{Milanovic1}, \citeyear{Milanovic1}, \citeyear{Milanovic}, \citeauthor{AP}, \citeyear{AP}, \citeauthor{APS}, \citeyear{APS}, \citeauthor{Toda}, \citeyear{Toda}, \citeauthor{IIW}, \citeyear{IIW}, \citeauthor{GabaixIneq}, \citeyear{GabaixIneq}, \citeauthor{Ibr}, \citeyear{Ibr},  \citeauthor{TodaWang}, \citeyear{TodaWang}, and references therein).\footnote{As is well-known, heavy-tailedness and power law distributons are also exhibited by many other key variables in economics and finance, including financial returns, foreign exchange rates, insurance risks and losses from natural disasters, to name a few (see, among others, the reviews in \cite{EKM}, \cite{Gabaix1}, \cite{IIW}, \cite{MFE}, and references therein.} 

Typical empirical results are $\zeta\in (1.5, 3)$ for income, and $\zeta\approx 1.5$ for wealth. Thus, the variance is infinite for wealth and may be infinite for income.\footnote{Importantly, the value of the tail index $\zeta$ in power law income or wealth distributions (\ref{power1}) may be regarded as a measure of upper tail inequality (that is, among the rich), with smaller values of the  tail index corresponding to larger inequality in the upper tails. This may be motivated by the fact that, in the case of Pareto distributions with $\zeta>1$ for income or wealth, where (\ref{power1}) holds exactly for all values $x$ greater than a certain threshold $x_m,$ the Gini coefficient of inequality over the whole income/wealth distribution is equal to $1/ (2\zeta -1)$ and is thus decreasing in $\zeta$ (see also the discussion in \citeauthor{Atkinson}, \citeyear{Atkinson}, \citeauthor{GabaixIneq}, \citeyear{GabaixIneq}, that focuses on the analysis and estimation of the top income inequality measure $\eta=1/\zeta,$ \citeauthor{Clara}, \citeyear{Clara}, and \citeauthor{Ibr}, \citeyear{Ibr}). } 

More generally, the tail index parameter $\zeta$ of power law distributions (\ref{power1}) characterize the heaviness (the rate of decay) of its tails, with smaller values of $\zeta$ corresponding to more pronounced heavy-tailedness in the distributions, and vice versa. The tail index $\zeta$ governs the likelihood of observing outliers and extreme values of $X,$ e.g., very high income/wealth levels in the case of income and wealth distributions. It is further important as it governs existence of moments of the r.v. $X>0,$ with the moment $EX^p$ of order $p>0$ of $X$ being finite if and only if $\zeta>p.$ In particular, the second moment $EX^2$ of the r.v. $X$ is finite and its variance $Var(X)$ is defined if and only if $\zeta>2,$ and the first moment - the mean $EX$- of $X$ is finite if and only if $\zeta>1.$ 


Applicability of commonly used approaches to inference on inequality measures based on asymptotic normality becomes problematic under heavy-tailedness, heterogeneity and correlation in the data. For instance, sample inequality measures - estimators of measures of inequality like sample Gini coefficient - converge to non-Gaussian limits given by 
stable random variables (r.v.'s) under sufficiently pronounced heavy-tailedness with infinite second moments and variances (see \cite{Taleb} and the discussion in Appendix B).\footnote{As is well-known, finiteness of variances for variables dealt with, such as economic and financial indicators like financial returns and exchange rates, is crucial for applicability of standard statistical and econometric approaches, including regression and least squares methods. Similarly, the problem of potentially infinite fourth moments of (economic and financial) variables and time series dealt with needs to be taken into account in applications of
autocorrelation-based methods and related inference procedures in their analysis (see, among others, the discussion in \citeauthor{Granger},  \citeyear{Granger}, \citeauthor{EKM},  \citeyear{EKM}, \citeauthor{cont2001empirical}, \citeyear{cont2001empirical}, Ch. 1 in \citeauthor{IIW}, \citeyear{IIW}, and references therein).} 


Provided normal convergence for sample inequality measures holds, asymptotic methods based on it often have poor finite sample properties under the problems of extreme values, outliers and heavy-tailedness in data.\footnote{More generally, poor finite sample properties are often observed for asymptotic methods based on normal convergence of estimators and consistent estimation of their limiting variances under heterogeneity and dependence in observations (e.g., inference approaches based on heteroskedasticity and autocorrelation consistent - HAC - and clustered standard errors, especially with data with pronounced autocorrelation, dependence and heterogeneity, see, among others, \citeauthor{Andrews}, \citeyear{Andrews}, 
\citeauthor{DL}, \citeyear{DL}, the discussion in \citeauthor{Ph}, \citeyear{Ph}, \citeauthor{IM2}, \citeyear{IM2, IM1}, \citeauthor{canay}, \citeyear{canay}, 
\citeauthor{esarey}, \citeyear{esarey}, and references therein).} Similar problems are also observed for bootstrap methods (see \citeauthor{CF}, \citeyear{CF}, \citeauthor{DF}, \citeyear{DF}). Bootstrap methods are also known to fail in heavy-tailed infinite variance settings (see the discussion in Section 5 in \citeauthor{DF}, \citeyear{DF}, and references therein).

The problems with inference on inequality measures are discussed in detail in \cite{Dufour, Dufour1}. These works also emphasize that reliable methods remain scarce for both the one-sample problem of inference on a single inequality index and the two-sample problem of testing for equality of and inference on the difference between two inequality indices. As discussed in \cite{Dufour, Dufour1}, the latter problem is much more challenging than the former (see also \citeauthor{IM1}, \citeyear{IM1}, for the discussion and the results on robust inference on equality of and the difference between two general parameters of interest under heterogeneity and dependence). \cite{Dufour} propose permutation tests for the hypothesis of equality of two inequality measures from independent samples which outperforms other asymptotic and bootstrap methods available in the literature (see also \citeauthor{canay}, \citeyear{canay}, for permutation tests of equality of two general parameters of interest under heterogeneity and clustered dependence). As discussed in \cite{Dufour1}, the latter tests for the two-sample problem for inequality indices are limited to testing the equality of two inequality measures. In particular, they do not provide a way of making inference on a possibly non-zero difference between the two measures considered nor building a confidence interval for the difference.

\cite{Dufour1} propose Fieller-type methods for inference on the generalized entropy (GE) class of inequality measures. Among other results, the authors develop approaches to testing and construction of confidence intervals for any possibly non-zero difference between the inequality measures that can be used under independent samples of i.i.d. observations with possibly unequal sizes and equal-sized samples of i.i.d. observation with arbitrary dependence between the samples.

This paper focuses on applications of recently developed $t-$statistic approaches (see \citeauthor{IM2}, \citeyear{IM2, IM1}, and also Ch. 3 in \citeauthor{IIW}, \citeyear{IIW}) in robust inference on income and wealth inequality measures under the problems of heterogeneity, heavy-tailedness and possible dependence in observations. Following the approaches, in particular, a robust large sample test on equality or a non-zero difference of two parameters of interest (e.g., a test of equality of inequality measures in two regions or countries considered) is conducted as follows: The data in the two samples dealt with is partitioned into fixed numbers $q_1, q_2\ge 2$ (e.g., $q_1=q_2=2, 4, 8$) of groups, the parameters (inequality measures dealt with) are estimated for each group, and inference is based on a standard two-sample $t-$test with the resulting $q_1, q_2$ group estimators (see the next section). As follows from the results in \cite{IM2, IM1}, robust $t-$statistic approaches result in valid inference under general conditions that group estimators of parameters of interest (e.g., inequality measures) considered weakly converge, at an arbitrary rate, to independent normal or scale mixtures of normal r.v.'s. These conditions are typically satisfied in empirical applications even under pronounced heavy-tailedness, heterogeneity and possible dependence in observations.\footnote{In particular, as discussed in \cite{IM2} and Ch. 3 in \cite{IIW}, the asymptotic Gaussianity of group estimators of the parameters of interest typically follows from the same reasoning and holds under the same conditions as the asymptotic Gaussianity of their full-sample estimators.} The approaches proposed in the paper complement and compare favorably with other inference methods available in the literature, including computationally expensive bootstrap procedures and permutation-based inference methods. Importantly, the approaches proposed in the paper can be used in testing and construction of confidence intervals for any possibly non-zero difference between inequality measures under the problems of heterogeneity, heavy-tailedness and possible dependence in the data. 

One should also emphasize wider range of applicability of $t-$statistic approaches to inference on inequality measures proposed in the paper as compared to other inference methods available in the literature, including those considered in \cite{Dufour, Dufour1}. The inference approaches can be used in the case where observations (e.g., on income or wealth levels) in each of the samples considered are \emph{dependent} among themselves - for instance, due to spatial or clustered dependence (see \cite{Conley} and \cite{Bhat} for a review of settings and methods of inference under spatial and clustered dependence, including complex stratified and clustered household surveys), common shocks affecting them (see \cite{AndrewsCommon} and \cite{Hwang} for a review of and inference using data with common shock dependence), or, in the case of time series or panel data on income or wealth levels, due to autocorrelation and dependence in observations over time. Further, in the case of testing for equality of inequality measures or inference on their difference in two populations using two samples of possibly dependent observations, as above, the $t-$statistic inference approaches may be used under \emph{an arbitrary} dependence \emph{between} the samples as well as under possibly \emph{unequal} sample sizes. 


Application of the robust inference approaches is illustrated by an empirical analysis of income inequality measures and their comparisons across different regions in Russia.

 The paper is organized as follows. Section \ref{tstat} describes the robust $t-$statistic approaches to inference on inequality measures analyzed in the paper and discusses the conditions for their validity. Section \ref{num} provides numerical results on finite sample performance of the robust inference approaches dealt with and their comparisons with other inference methods in the literature, with a particular focus on testing equality of two inequality measures and inference on the difference between two inequality indices in Section \ref{twosample}. Section \ref{emp} presents empirical applications of the robust $t-$statistic approaches in the analysis of income inequality in Russia and comparisons of inequality measures across Russian regions. Section \ref{conclude} makes some concluding remarks and discusses some suggestions for future research.  Appendix A provides tables on the numerical and empirical results in the paper. 
 Appendix B provides a review of the definitions and asymptotic properties of the Gini, Theil and Generalized Entropy measures referred to in the paper and a discussion of applicability of the robust approaches dealt with in inference on the measures. 

\section{Methodology: Robust \textit{t}-statistic approaches to inference on inequality measures} \label{tstat}



We focus on inference on inequality measures using the \textit{t-}statistic approaches to robust inference under heterogeneity, heavy-tailedness and dependence of largely unknown form recently developed in \cite{IM2, IM1}. \cite{IM2} provide an approach to robust inference on an arbitrary single parameter of interest. \cite{IM1} provide approaches to 
robust testing of equality of two arbitrary parameters of interest and to robust inference on the difference of the parameters.

We refer to, among others, Section 13.F in \cite{CF}, \cite{DF}, \cite{MO}, \cite{Ibr}, \cite{Dufour} and \cite{Dufour1} for definitions of the most widely inequality measures, including Gini, Generalized Entropy and Theil indices, and their values for different income distributions, including  empirically relevant heavy-tailed Pareto, double Pareto and Singh-Maddala distributions (see the next section).

In the context of a one-sample inference on a single (income or wealth) inequality (e.g., a Theil, Generalized entropy - GE - or Gini index) the robust $t-$statistic approaches are implemented as follows.


Throughout the paper, we denote by $T_k$ a r.v. that has a Student-$t$ distribution with $k\ge 1$ degrees of freedom. Further, for $q\ge 2$ and $0<\alpha<1,$ by $cv_{q, \alpha}$ we denote the $(1-\alpha/2)-$quantile of the Student-$t$ distribution with $q-1$ degrees of freedom: $P(\left|T_{q-1}\right|>t_{\alpha }$)=\textit{ $\alpha$}.


Consider the one-sample problem of testing a hypothesis on or constructing a confidence interval for an inequality measure $\mathcal{L}.$ 
Following the $t-$statistic robust inference approaches in \cite{IM2}, a (large) sample \textit{I}${}_{1}$\textit{, I}${}_{2}$\textit{,..., I${}_{N}$ } of observations on income or wealth levels $I,$ is partitioned into a fixed number $q\ge 2$ (e.g., $q=2, 4$ or 8) of groups, and the income inequality measure $\mathcal{L}$\textit{ } is estimated using the data for each group thus resulting in $q$ group empirical income inequality measures ${\widehat{\mathcal{L}}}_j$, $j = 1,...,q.$
The robust test of the null hypothesis  $H_0: \mathcal{L}={\mathcal{L}}_0$ against the two-sided alternative $H_a: \mathcal{L}\neq {\mathcal{L}}_0$ 
is based on the usual $t-$statistic $t^{I}_{\mathcal{L}}$ in the $q$ group empirical inequality measures ${\widehat{\mathcal{L}}}_j$, $j = 1,...,q:$ \begin{eqnarray} \label{tstat1} t_{\mathcal{L}}=\sqrt{q}\frac{\overline{\widehat{\mathcal{L}}}-{\mathcal{L}}_0}{s_{\widehat{\mathcal{L}}}} \end{eqnarray} with  $\overline{\widehat{\mathcal{L}}}=\frac{\sum^q_{j=1}{{\widehat{\mathcal{L}}}_j}}{q}$\textit{ }and $s^2_{\widehat{\mathcal{L}}}=\frac{\sum^q_{j=1}{{\left({\widehat{\mathcal{L}}}_j-\overline{\widehat{\mathcal{L}}}\right)}^2}}{q-1}$  ${\widehat{\mathcal{L}}}_j.$ 
The above null hypothesis $H_0$ is rejected in favor of the alternative $H_a$ at level $\alpha\le 0.83$ (e.g., at the usual significance level $\alpha=0.05$) if the absolute value $|t_{\mathcal{L}}|$ of the $t-$statistic in group estimates ${\widehat{\mathcal{L}}}_j$ exceeds the $(1-\alpha/2)-$quantile of the standard Student-$t$ distribution with $q-1$ degrees of freedom: $|t_{\mathcal{L}}|>cv_{q, \alpha}.$ 
The test of $H_0$ against $H_a$ of level $\alpha\le 0.1$ is conducted in the same way if $2\le q\le 14.$ Using the results in \cite{bakirov2006student} and \cite{IM2}, one can further calculate the $p-$values of the above $t-$statistic robust tests in the case of an arbitrary number $q$ of groups thus enabling conducting robust tests on the inequality measure $\mathcal{L}$ of an arbitrary level.\footnote{One-sided tests are conducted in a similar way; one may note that quantiles of Student-$t$ distributions with $q-1$ degrees of freedom can also be used in one-sided tests of level $\alpha\le 0.1$ if $q\in \{2, 3\}.$}

By implication, for all $\alpha\le 0.83$ (and all $\alpha\le 0.1$ for $2\le q\le 14$) a confidence interval for the inequality measure $\mathcal{L}$
with asymptotic coverage of at least $1-\alpha$ may be constructed as ${\widehat{\mathcal{L}}}_j\pm cv_{q, \alpha} s_{\widehat{\mathcal{L}}}.$ 
For instance, the 95\% confidence interval for $\mathcal{L}$\textit{ }is given by $(\overline{\widehat{\mathcal{L}}}-cv_{q, 0.05}s_{\widehat{\mathcal{L}}}\mathrm{,\ \ }\overline{\widehat{\mathcal{L}}}+cv_{q, 0.05}s_{\widehat{\mathcal{L}}}$)\textit{, }where \textit{cv}${}_{q, 0.05}$ is the 0.975-quantile of the Student-\textit{t} distribution with \textit{q$-$}1 degrees of freedom: $P(\left|T_{q-1}\right|>cv_{q, 0.05}$)=0.05. 
As follows from \cite{IM2}, the above approach results in asymptotically valid inference under the assumption that the group empirical income inequality measures ${\widehat{\mathcal{L}}}_j$, $j=1,..., q,$ are asymptotically independent, unbiased and Gaussian of possibly different variances.  

The asymptotic validity of the \textit{t-}statistic based inference approach continues to hold even when the group estimators ${\widehat{\mathcal{L}}}_j$ of $\mathcal{L}$\textit{ }converge (at an arbitrary rate) to independent but potentially heterogeneous scale mixtures of normal r.v.'s, such as heavy-tailed stable symmetric r.v.'s. It also holds under convergence of the group estimators to conditionally normal r.v.'s which are unconditionally dependent through their second moments or have a common shock-type dependence (see \citeauthor{AndrewsCommon}, \citeyear{AndrewsCommon}, for inference methods under common shock dependence structures, and \citeauthor{Hwang}, \citeyear{Hwang}, for applications of $t-$statistic robust inference approaches in such settings).\footnote{Justification of asymptotic validity of the robust $t-$statistic inference approaches in \cite{IM2} is based on a small sample result in \cite{bakirov2006student} that implies validity of the standard $t-$test on the mean under independent heterogeneous normal observations. Justification of asymptotic validity of the approaches in inference on equality of two parameters in \cite{IM1} is based on the analogues of the above small sample result for two-sample $t-$tests and Behrens-Fisher problem obtained therein.} This implies that the \textit{t-}statistic based robust inference on $\mathcal{L}$ can thus be applied under extremes and outliers in observations generated by heavy-tailedness with infinite variances and, among others, dependence structures that include models with multiplicative common shocks (see \citeauthor{Ibr1}, \citeyear{Ibr1}, \citeyear{Ibr2}). The $t-$statistic based approaches do not require at all estimation of limiting variances of estimators of interest, in contrast to inference methods based on consistent, e.g., HAC or clustered, standard errors (see Section \ref{intro}). The numerical analysis in \cite{IM0, IM1} and Section 3 in \cite{IIW} indicates favorable finite sample performance of the $t-$statistic based robust inference approaches in inference on models with time series, panel, clustered and spatially correlated data. See also \cite{esarey} for a detailed numerical analysis of finite sample performance of different inference procedures, including $t-$statistic and related approaches,  under small number of clusters of dependent data  and their software (STATA and R) implementation.

The above conditions for asymptotic validity of $t-$statistic approaches to robust inference are typically satisfied in applications, under the appropriate choice of the groups implying asymptotic unbiasedness and independence of group estimators of parameters of interest - inequality measures considered (see below). Namely, the asymptotic Gaussianity (or other weak convergence results, e.g., convergence to heavy-tailed scale mixtures of Gaussian distributions) of group estimators - group empirical inequality measures - ${\widehat{\mathcal{L}}}_j$ typically follows from the same reasoning and holds under the same conditions as the asymptotic Gaussianity (or other relevant asymptotics) of the full-sample estimator - full-sample empirical inequality measure - ${\widehat{\mathcal{L}}}.$

Concerning the choice of the groups, the condition that group estimators of parameters of interest - inequality measures dealt with - should be asymptotically unbiased (and independent) places natural - again typically satisfied in applications - restrictions on formation of groups in applications of $t-$statistic approaches in the context of inference on inequality indices and their comparisons (see also discussion of general $t-$statistic inference approaches in \citeauthor{IM2}, \citeyear{IM2}). For instance, in the problem of inference on a single inequality measure in the whole country, e.g., Russia, using household income surveys with random samples of households in the country and its regions and thus i.i.d data on income levels, groups cannot be chose to be the country regions. This is because each of the group estimators - group inequality measures - will estimate the inequality index considered in the corresponding region but not in the whole country and unbiasedness of the group estimators with the mean asymptotically equal to the country's inequality index of interest will not hold. The ``between-region'' component of inequality in the whole country would be missed out by the group estimators.

On the other hand, in the problem of testing equality of or inference on the difference between inequality indices in two regions of a country, e.g., Russia as in the empirical application in Section \ref{emp} in this paper, using household income surveys with i.i.d. data on household income levels in the regions considered, the groups in applications of two-sample $t-$statistic approaches can be formed just by taking subsequent observations on incomes in the two samples of i.i.d. income data in the regions (similar to applications of the approaches with time series data, see \citeauthor{IM2}, \citeyear{IM2}). Namely, in the case of inference on equality of inequality indices of interest in two regions using the random samples $I_1, I_2, ..., I_{N_1},$ $Y_1, Y_2, ..., Y_{N_2}$ of (i.i.d.) income levels in them, the $q_1, q_2$ groups in applications of two-sample $t-$statistic approaches based on $\tilde{t}_{\mathcal{L}}$ in (\ref{Gen2}) can be taken to be the groups $\{I_k, (i-1)N_1/q_1<k\le iN_1/q_1\},$ $\{Y_l, (j-1)N_2/q_2<l\le jN_2/q_2\},$ $i=1, ..., q_1,$ $j=1, ..., q_2,$ of subsequent observations on household incomes in the samples considered. The groups of subsequent observations in the two samples in applications of $t-$statistic approaches based on $\tilde{\tilde{t}}_{\mathcal{L}}$ in (\ref{GenDiff}) with $q_1=q_2=q$ are formed in a similar way. With the above simple choice of groups, asymptotic unbiasedness and independence of group estimators - group inequality measures - holds due to i.i.d.ness of data in the random samples considered.

Let us now turn to testing that the values $L_1$ and $L_2$ of an inequality measure $\mathcal{L}$ are equal in two populations (e.g., for income distributions in two regions of a country) and to inference on the difference $d=L_1-L_2$ between the two inequality indices using the (large) samples $I_1, I_2, ..., I_{N_1}$ and $Y_1, Y_2, ..., Y_{N_2}$ on income or wealth levels in the populations. We first assume that the two samples are independent. Following the $t-$statistic approaches to robust inference on two parameters of interest in \cite{IM1}, each of the two samples $I_1, I_2, ..., I_{N_1}$ and $Y_1, Y_2, ..., Y_{N_2}$ is partitioned into fixed numbers $q_1, q_2\ge 2$ (e.g., $q_1, q_2=2, 4$ or 8) groups, respectively, and the income inequality measure $\mathcal{L}$\textit{ } is estimated using the data for each of the groups in the two samples. This thus results in $q_1+q_2$ group empirical income inequality measures ${\widehat{\mathcal{L}}}^{I}_1,$ ..., ${\widehat{\mathcal{L}}}^{I}_{q_1},$ and ${\widehat{\mathcal{L}}}^{Y}_1$, ...,  ${\widehat{\mathcal{L}}}^{Y}_{q_2}.$ 
The robust test of the null hypothesis  $H_0: L_1-L_2=d_0$ (e.g., with $d_0=0,$ the test of the hypothesis $H_0: L_1=L_2$ of equality of the values $L_1, L_2$ of the inequality index $\mathcal{L}$ in the two populations considered) against the two-sided alternative $H_a: L_1-L_2\neq d_0$ (resp., with $d_0=0,$ against the two-sample alternative $H_a: L_1\neq L_2$) is based on the usual two-sample $t-$statistic $\tilde{t}_{\mathcal{L}}$ in the $q_1+q_2$ group empirical inequality measures ${\widehat{\mathcal{L}}}^{I}_j$,  ${\widehat{\mathcal{L}}}^{Y}_k$, $j = 1,...,q_1,$ $k = 1,...,q_2:$ \begin{eqnarray} \label{Gen2} \tilde{t}_{\mathcal{L}}=\frac{
\overline{\widehat{\mathcal{L}}^I}-\overline{\widehat{\mathcal{L}}^Y}-d_0}{\sqrt{s^2_{\widehat{\mathcal{L}}^I}/q_1+s^2_{\widehat{\mathcal{L}}^Y}/q_2}}\end{eqnarray} with  $$\overline{\widehat{\mathcal{L}}^I}=\frac{1}{q_1}\sum^{q_1}_{j=1}{{\widehat{\mathcal{L}}}^I_j}, \overline{\widehat{\mathcal{L}}^Y}=\frac{1}{q_2}\sum^{q_2}_{k=1}{{\widehat{\mathcal{L}}}^Y_k},$$ $$ s^2_{\widehat{\mathcal{L}}^I}=\frac{1}{q_1-1}\sum^{q_1}_{j=1}{\left({\widehat{\mathcal{L}}}^I_j-\overline{\widehat{\mathcal{L}}^I}\right)^2}, s^2_{\widehat{\mathcal{L}}^Y}=\frac{1}{q_2-1}\sum^{q_2}_{k=1}{{\left({\widehat{\mathcal{L}}}^Y_k-\overline{\widehat{\mathcal{L}}^Y}\right)}^2}.$$

For the number of groups $q_1, q_2\le 14$ the above null hypothesis $H_0: L_1-L_2=d_0$ is rejected in favor of the alternative $H_a: L_1-L_2\neq d_0$ (resp., with $d_0=0,$  the null hypothesis $H_0: L_1=L_2$ of equality of the inequality measures values in the populations is rejected in favor of the alternative $H_a: L_1\neq L_2$) at level  $\alpha \in \{0.001, 0.002, ... , 0.099, 0.10\}$ (e.g., at the usual significance levels $\alpha=0.01, 0.05$ and 0.1) if the absolute value $|\tilde{t}_{\mathcal{L}}|$ of the two-sample $t-$statistic in group empirical inequality measures ${\widehat{\mathcal{L}}}^{I}_j$,  ${\widehat{\mathcal{L}}}^{Y}_k$, $j = 1,...,q_1,$ $k = 1,...,q_2,$ exceeds the $(1-\alpha/2)-$quantile of the standard Student-$t$ distribution with $q-1$ degrees of freedom, where $q=\min(q_1, q_2):$  $|\tilde{t}_{\mathcal{L}}|>cv_{q, \alpha}=cv_{\min(q_1, q_2), \alpha}.$ \footnote{As in the one-sample case, one-sided tests are conducted in a similar way.}\footnote{As follows from the analysis in \cite{IM1}, the described tests may also be used for all $q_1, q_2\le 50$ if $\alpha \in \{0.001, 0.002, ... , 0.083\},$ e.g., for the usual critical values $\alpha=0.01, 0.05.$} 

One further obtains that, for $\alpha=0.01, 0.05, 0.1$ and the number of groups $q_1, q_2\le 14,$ $\min(q_1, q_2)=q,$ a confidence interval for the difference $d_0=L_1-L_2$ between the values of the inequality measure $\mathcal{L}$ in two populations
with asymptotic coverage of at least $1-\alpha$ may be constructed as \\ ${\widehat{\mathcal{L}}}^I-{\widehat{\mathcal{L}}}^Y\pm cv_{q, \alpha}\sqrt{s^2_{\widehat{\mathcal{L}}^I}/q_1+s^2_{\widehat{\mathcal{L}}^Y}/q_2}.$  For instance, the 95\% confidence interval for $\mathcal{L}$\textit{ }is given by 
${\widehat{\mathcal{L}}}^I-{\widehat{\mathcal{L}}}^Y\pm cv_{q, 0.05}\sqrt{s^2_{\widehat{\mathcal{L}}^I}/q_1+s^2_{\widehat{\mathcal{L}}^Y}/q_2},$ where $cv_{q, 0.05}$ is the 0.975-quantile of the Student-\textit{t} distribution with $\min(q_1, q_2)-1$ degrees of freedom: $P(\left|T_{\min(q_1, q_2)-1}\right|>cv_{q, 0.05}$)=0.05.

As follows from \cite{IM1}, the two-sample $t-$statistic approach is asymptotically valid under the assumption - as above, typically satisfied in applications - that the group empirical income inequality measures ${\widehat{\mathcal{L}}}^I_j$, $j=1,..., q_1,$ ${\widehat{\mathcal{L}}}^Y_k$, $k=1,..., q_2,$ are asymptotically independent, unbiased and Gaussian of possibly different variances (or converge at an arbitrary rate to independent but potentially heterogeneous scale mixtures of normal r.v.'s, such as heavy-tailed stable symmetric r.v.'s).

Let us now consider the problem of testing for equality of the values $L_1$ and $L_2$ of an inequality measure $\mathcal{L}$ and to inference on the difference $d=L_1-L_2$ between the inequality indices in two populations using income or wealth level samples $I_1, I_2, ..., I_{N_1}$ and $Y_1, Y_2, ..., Y_{N_2}$ of \emph{possibly unequal sizes $N_1, N_2$}  that may exhibit an \emph{arbitrary dependence between them}. Suppose that the samples are divided into an equal number $q_1=q_2=q\ge 2$  (e.g., $q_1, q_2=2, 4$ or 8) of groups, and the sample inequality measures - estimates of the inequality index $\mathcal{L}$ of interest - are calculated using the data for each of the $2q$ groups in the two samples. One thus has the 
group empirical income inequality measures ${\widehat{\mathcal{L}}}^{I}_1,$ ..., ${\widehat{\mathcal{L}}}^{I}_{q},$ and ${\widehat{\mathcal{L}}}^{Y}_1$, ...,  ${\widehat{\mathcal{L}}}^{Y}_{q}.$ 
The robust test of the null hypothesis  $H_0: L_1-L_2=d_0$ (e.g., with $d_0=0,$ the test of the hypothesis $H_0: L_1=L_2$ of equality of the values $L_1, L_2$ of the inequality index $\mathcal{L}$ in the two populations) against the two-sided alternative $H_a: L_1-L_2\neq d_0$ (resp., with $d_0=0,$ against the two-sample alternative $H_a: L_1\neq L_2$) may be based on the one-sample $t-$statistic $\tilde{\tilde{t}}_{\mathcal{L}}$ in the $q$ differences ${\widehat{\mathcal{L}}}^{I}_j-{\widehat{\mathcal{L}}}^{Y}_j,$ $j = 1,...,q,$ of the calculated group empirical inequality measures: \begin{eqnarray} \label{GenDiff} \tilde{\tilde{t}}_{\mathcal{L}}=\sqrt{q}\frac{
\overline{\widehat{\mathcal{L}}^I}-\overline{\widehat{\mathcal{L}}^Y}-d_0}{s_{\widehat{\mathcal{L}}^{I-Y}}}\end{eqnarray} with  $$\overline{\widehat{\mathcal{L}}^I}=\frac{1}{q}\sum^{q}_{j=1}{{\widehat{\mathcal{L}}}^I_j}, \overline{\widehat{\mathcal{L}}^Y}=\frac{1}{q}\sum^{q}_{j=1}{{\widehat{\mathcal{L}}}^Y_j},$$ $$ s^2_{\widehat{\mathcal{L}}^{I-Y}}=\frac{1}{q-1}\sum^{q}_{j=1}{\left(({\widehat{\mathcal{L}}}^I_j-{\widehat{\mathcal{L}}}^Y_j)-(\overline{\widehat{\mathcal{L}}^I}-\overline{\widehat{\mathcal{L}}^Y})\right)^2}.$$

As in the case of $t-$statistic inference on one parameter, for any $\alpha\le 0.83$ (any $\alpha\le 0.1$ for $2\le q\le 14$), the null hypothesis $H_0: L_1-L_2=d_0$ (for $d_0=0,$ the null hypothesis $H_0: L_1=L_2$ of equality of the values $L_1, L_2$ of the inequality index $\mathcal{L}$ in two populations) is rejected in favor of the two-sided alternative $H_a: L_1-L_2\neq d_0$ (resp., for $d_0=0,$ in favor of the alternative $H_a: L_1\neq L_2$)  at level $\alpha$ if the absolute value $|\tilde{\tilde{t}}_{\mathcal{L}}|$ of the $t-$statistic in the differences ${\widehat{\mathcal{L}}}^I_j-{\widehat{\mathcal{L}}}^Y_j,$ $j=1, ..., q$ of group sample inequality measures exceeds the $(1-\alpha/2)-$quantile of the standard Student-$t$ distribution with $q-1$ degrees of freedom: $|\tilde{\tilde{t}}_{\mathcal{L}}|>cv_{q, \alpha}.$ Further, as in the case of $t-$statistic inference on a single inequality measure, the $p-$values of the above tests can be calculated in the case of an arbitrary number $q=q_1=q_2$ of groups thus enabling conducting robust tests of an arbitrary level.

For all $\alpha\le 0.83$ (and all $\alpha\le 0.1$ for $2\le q\le 14$)  a confidence interval for the difference $d_0=L_1-L_2$ between the values of the inequality measure $\mathcal{L}$ in two populations
with asymptotic coverage of at least $1-\alpha$ may be constructed as ${\widehat{\mathcal{L}}}^I_j-{\widehat{\mathcal{L}}}^Y_j\pm cv_{q, \alpha} s_{\widehat{\mathcal{L}}^{I-Y}}.$ For instance, the 95\% confidence interval for $\mathcal{L}$\textit{ }is given by $({\widehat{\mathcal{L}}}^I_j-{\widehat{\mathcal{L}}}^Y_j- cv_{q, 0.05} s_{\widehat{\mathcal{L}}^{I-Y}}, {\widehat{\mathcal{L}}}^I_j+{\widehat{\mathcal{L}}}^Y_j- cv_{q, 0.05} s_{\widehat{\mathcal{L}}^{I-Y}}),$ where \textit{cv}${}_{q, 0.05}$ is the 0.975-quantile of the Student-\textit{t} distribution with \textit{q$-$}1 degrees of freedom: $P(\left|T_{q-1}\right|>cv_{q, 0.05}$)=0.05.

As above, the $t-$statistic approaches to robust inference based on (\ref{GenDiff}) are asymptotically valid under the assumption that the group empirical income inequality measures ${\widehat{\mathcal{L}}}^I_j$,  ${\widehat{\mathcal{L}}}^Y_j$, $j=1,..., q,$ are asymptotically independent across $j$, unbiased and Gaussian of possibly different variances (or have limiting scale mixtures of Gaussian distributions).

\section{Finite sample performance} \label{num}

In this section, we provide numerical results on finite sample properties of the asymptotic, robust $t-$statistic, bootstrap  and permutation approaches to inference and tests on inequality measures. The results are provided for inference on Theil and Gini inequality, similar to the numerical analysis in \cite{CF} and \cite{Dufour}. 


We first present the results for the one-sample problem of inference on a single inequality measure in Section \ref{onesample}. 

Then, in Section \ref{twosample}, the numerical results are provided for the two-sample problem of tests on equality of two inequality measures and inference on the difference between two inequality indices.



As in \cite{Dufour}, the numerical analysis of finite-sample performance of different approaches to inference on inequality measure(s) is based on simulations from Singh-Maddala distributions that were reported to provide a good fit to real-world income distributions in various countries (see the discussion in \citeauthor{CF}, \citeyear{CF}, \citeauthor{DF}, \citeyear{DF}, \citeauthor{Dufour}, \citeyear{Dufour}, and references therein). The cdf of the Singh-Maddala distribution with the scale parameter $b>0$ and the shape parameters $a, c>0$ is given by
\begin{eqnarray} \label{SM} F(x)=1-\left[1+\left(\frac{x} {b}\right)^a\right]^{-c}, x>0,\end{eqnarray}
(see the above references). Similar to \cite{Dufour}, the Singh-Madalla distribution with parameters $a, b, c>0$ is denoted by $SM(a, b, c)$ in what follows. It is easy to see that the cdf $F(x)$ of the Singkh-Maddala distribution $SM(a, b, c)$ satisfies $F(x)\sim c\left(\frac{x} {b}\right)^a$ as $x\rightarrow 0,$ and $1-F(x)\sim \left(\frac{x}{b}\right)^{-ac}$ as $x\rightarrow \infty.$
\footnote{As usual, we write $f(x)\sim g(x)$ as $x\rightarrow x_0$ or $x\rightarrow \infty$ for two positive functions $f(x)$ and $g(x)$ if $f(x)/g(x)\rightarrow 1$ as $x\rightarrow x_0$ or $x\rightarrow \infty$.} Therefore, the Singh-Maddala distribution has the (double) power law or (double) Pareto behavior in the lower tails - for small income levels - and the upper tails - for high incomes (see \citeauthor{Toda}, \citeyear{Toda}, for the analysis of double Pareto and related distributions for income). 

In particular, the Singh-Maddala distributions belong to the class of distributions with heavy power law tails, so that for for large $x>0$ and r.v.'s (income or wealth levels) $X>0$ with the Singh-Maddala distribution $SM(a, b, c)$ follows power law (\ref{power1}) with the tail index $\zeta=ac.$ 

Following \cite{Dufour}, in the numerical experiments, we use the parameter values $a_0=2.8,$ $b_0=100^{-1/2.8},$ $c_0=1.7$ for the Singh-Maddala distribution, with the corresponding tail index $\zeta=a_0c_0=4.76,$ as a benchmark. The Theil index for this distribition equals to 0.1401151, and the Gini index equals to 0.2887138 (see \citeauthor{Dufour}, \citeyear{Dufour}). The Singh-Maddala distribution with the above values for the parameters was also used in \cite{CF} and \cite{DF} to demonstrate poor finite-sample performance of asymptotic and bootstrap inference approaches. 


Further, as in  \cite{Dufour}, in the numerical experiments, we consider several other Singh-Maddala distributions $SM(a, b_0, c)$ with the above scale parameter $b_0=100^{-1/2.8}$ for which the Theil inequality index and the Gini index are the same as in the case of the distribution $SM(a_0, b_0, c_0),$ and equal to 0.1401151 (the Theil index) and 0.2887138 (the Gini index). 

Following \cite{Dufour}, in simulations involving the Theil index, we take the parameters $(a, c)$ of the Singh-Maddala distributions $SM(a, b_0, c)$ equal to $(2.5, 2.502199),$ $ (2.6, 2.149747),$ $(2.7, 1.894309),$ $(2.8, 1.7),$ $(3.0, 1.4223847),$ $(3.2, 1.2320215),$ $(3.4, 1.0922125),$ $(3.8, 0.8984488),$ \\ $(4.8, 0.6366578)$ and $(5.8, 0.4996163).$ The corresponding tail indices $\zeta$ of these distributions equal to $\zeta=6.26, 5.59, 5.11, 4.76, 4.27, 3.94, 3.71, 3.41, 3.06, 2.9.$

In simulations involving the Gini index, we take, as in \cite{Dufour}, the parameters $(a, c)$ of the Singh-Maddala distributions $SM(a, b_0, c)$ equal to (2.5,2.640350), (2.6,2.218091), (2.7,1.920967), (2.8,1.7), (3.0,1.3921126), (3.2,1.1866026), (3.4,1.0388049), (3.8,0.8387663), (4.8,0.5784599) and \\ (5.8,0.4473111). The corresponding tail indices $\zeta$ of these distributions equal to $\zeta=$6.6, 5.77, 5.19, 4.76, 4.18, 3.80, 3.53, 3.19, 2.78, 2.59.  

We note that the tail index $\zeta=2.78, 2.59, 2.9$ for the considered Singh-Maddala distributions lie in the interval (1.5, 3) as is typically the case for real-world income distributions, as discussed above.  Additionally, we also consider more heavy-tailed distributions $SM(a, b_0, c)$ with $(a, c)=(2,1.1),$ $(2,0.7)$ and $b_0=100^{-1/2.8}$. The corresponding tail indices $\zeta$ in power laws (\ref{power1}) for these distributions equal to $\zeta=2.2, 1.4.$






\subsection{One-sample problem: Inference on a single inequality measure} \label{onesample}
\subsubsection{Inference in one-sample problem: Finite-sample distributions} \label{fs}
We follow the notation in the previous sections that is largely similar to \cite{CF, DF} and \cite{Dufour}. In the numerical analysis in this section, $\mathcal{L}_0=\mathcal{L}(F)$ denotes the true value of the inequality measure $\mathcal{L}$ of interest (e.g., the Theil or Gini inequality index) for a population with the cdf $F$ considered. As before, $\hat{\mathcal{L}}=\hat{\mathcal{L}}(I_1, ..., I_N)$ denotes the full-sample estimator of $\mathcal{L}$ (the full-sample empirical inequality measure) calculated using a sample of observations $I_1, ..., I_N$ from the population. Further, as in Section \ref{tstat}, $\hat{\mathcal{L}}_j,$ $j=1, ..., q,$ denote the group estimators of $\mathcal{L}$ (group empirical inequality measures) in applications of $t-$statistic inference approaches.

Asymptotic approaches to inference on an inequality measure $\mathcal{L}$ are based on normal approximations to sample distributions of full-sample estimators $\hat{\mathcal{L}}$ of the measures (full-sample empirical inequality measures), more precisely, on standard normal approximations to sample distributions of (full-sample) $t-$statistics $S_{\hat{\mathcal{L}}}=(\hat{\mathcal{L}}-\mathcal{L}_0)/s.e._{\hat{\mathcal{L}}}$ calculated using these estimators, where $s.e._{\mathcal{L}}$ denotes the usual consistent standard error of $\hat{\mathcal{L}}$ (see the formulas for the empirical inequality measures considered and their standard errors in \citeauthor{CF}, \citeyear{CF}, \citeauthor{DF}, \citeyear{DF}, and \citeauthor{Dufour}, \citeyear{Dufour}). As discussed in Section \ref{tstat}, validity of $t-$statistic robust inference approaches requires weak convergence of group estimators $\hat{\mathcal{L}}_j,$ $j=1, ..., q,$ of the inequality measures $\mathcal{L}$ (without any Studentization/normalization of the group estimators by their standard errors in contrast to the $t-$statistics $S_{\hat{\mathcal{L}}}$ calculated using the full-sample estimators) to possibly heterogeneous Gaussian distributions (or scale mixtures of Gaussian distributions). Further (see the discussion in the introduction and the Section \ref{tstat}), asymptotic normality of group estimators $\hat{\mathcal{L}}_j$ holds under the same conditions as in the case of the full-sample estimators $\hat{\mathcal{L}}.$ 

We, therefore, begin the analysis with an assessment of finite-sample distributions of (full-sample) empirical inequality measures $\hat{\mathcal{L}}$  and the (full-sample) $t-$statistics $S_{\hat{\mathcal{L}}}$ calculated using them. We, in particular, focus on the assessment of closeness of the above finite-sample distributions to Gaussian ones.

We focus on comparisons of finite-sample distributions of the  (full-sample) $t-$statistics $S_{\hat{\mathcal{L}}}$ with those of the centered empirical inequality measures normalized by their true standard deviations, that is, of the statistics $Z_{\hat{\mathcal{L}}}=(\hat{\mathcal{L}}-\mathcal{L}_0)/\sigma_{\hat{\mathcal{L}}},$ where $\sigma_{\hat{\mathcal{L}}}^2=Var(\hat{\mathcal{L}}).$ The true values of the standard deviations $\sigma_{\hat{\mathcal{L}}}$ for the populations and sample sizes considered are obtained using direct simulations.

\begin{figure}[!h] 
  \centering
  \includegraphics[width=0.7\linewidth]{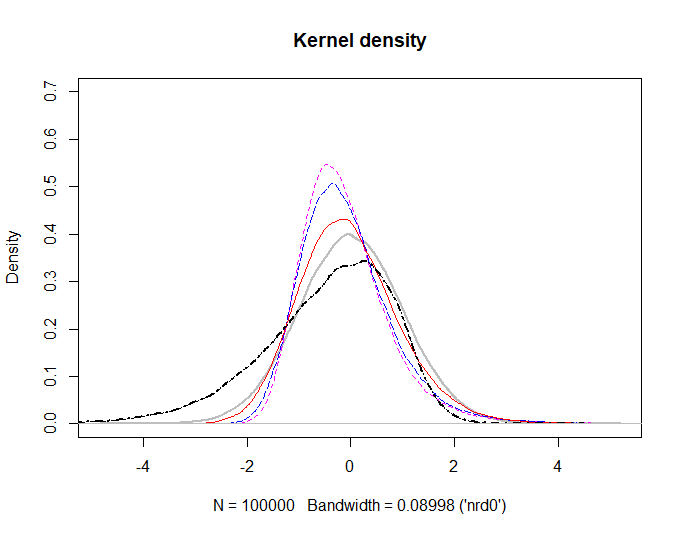}
  \caption{Kernel density functions for the statistics $Z_{\hat{\mathcal{L}}}$ and $S_{\hat{\mathcal{L}}}$ for the Theil index: Singh-Maddala distribution $SM(a_0, b_0, c_0)$ with $(a_0, c_0)=(2.8, 1.7)$ and $\zeta=4.76$. 
  Gaussian density: $\textcolor{gray}{\rule[0.25em]{2em}{1.6pt}\ }$;
Statistic $Z_{\hat{\mathcal{L}}}$, $N=50$: $\textcolor{magenta}{\rule[0.25em]{0.4em}{1.6pt} \ \rule[0.25em]{0.4em}{1.6pt}\ \rule[0.25em]{0.4em}{1.6pt} }$;
Statistic $Z_{\hat{\mathcal{L}}}$, $N=100$: $\textcolor{blue}{\rule[0.25em]{1em}{1.6pt} \ \rule[0.25em]{1em}{1.6pt}\ }$;  Statistic $Z_{\hat{\mathcal{L}}}$, $N=1000$: $\textcolor{red}{\rule[0.25em]{2em}{1.6pt} }$;
Statistic $S_{\hat{\mathcal{L}}}$, $N=100$: $\textcolor{black}{\rule[0.25em]{0.6em}{1.7pt} \ \mathbf{\cdot} \ \rule[0.25em]{0.6em}{1.7pt} \ }$
  }
  \label{kernel1}
\end{figure}
\begin{figure}[!h]
  \centering
  \includegraphics[width=0.7\linewidth]{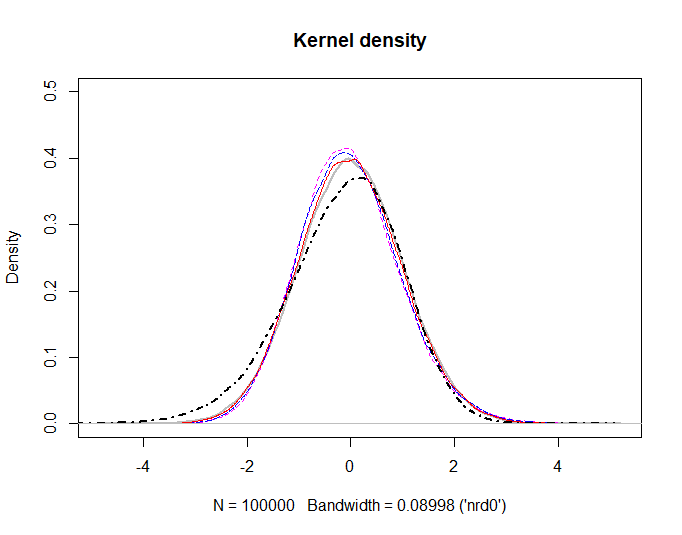}
  \caption{Kernel density functions for the statistics $Z_{\hat{\mathcal{L}}}$ and $S_{\hat{\mathcal{L}}}$ for the Gini index: Singh-Maddala distribution $SM(a, b_0, c_0)$ with $(a_0, c_0)=(2.8, 1.7)$ and $\zeta=4.76.$ 
  Gaussian density: $\textcolor{gray}{\rule[0.25em]{2em}{1.6pt}\ }$;
Statistic $Z_{\hat{\mathcal{L}}}$, $N=50$: $\textcolor{magenta}{\rule[0.25em]{0.4em}{1.6pt} \ \rule[0.25em]{0.4em}{1.6pt}\ \rule[0.25em]{0.4em}{1.6pt} }$;
Statistic $Z_{\hat{\mathcal{L}}}$, $N=100$: $\textcolor{blue}{\rule[0.25em]{1em}{1.6pt} \ \rule[0.25em]{1em}{1.6pt}\ }$;  Statistic $Z_{\hat{\mathcal{L}}}$, $N=1000$: $\textcolor{red}{\rule[0.25em]{2em}{1.6pt} }$;
Statistic $S_{\hat{\mathcal{L}}}$, $N=100$: $\textcolor{black}{\rule[0.25em]{0.6em}{1.7pt} \ \mathbf{\cdot} \ \rule[0.25em]{0.6em}{1.7pt} \ }$
  }
  \label{kernel2}
\end{figure}
\begin{figure}[!h]
  \centering
  \includegraphics[width=0.7\linewidth]{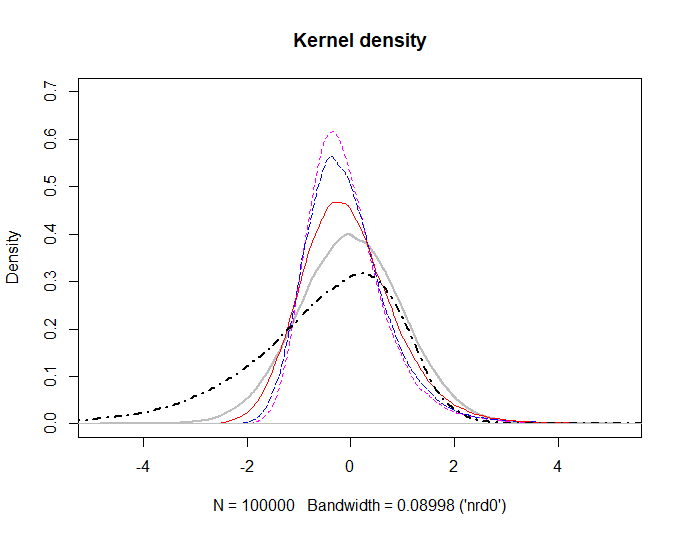}
  \caption{Kernel density functions for the statistics $Z_{\hat{\mathcal{L}}}$ and $S_{\hat{\mathcal{L}}}$ for the Gini index: Singh-Maddala distribution $SM(a, b_0, c)$ with $(a, c)=(5.8, 0.4473111)$ and  $\zeta=2.59$. 
  Gaussian density:$\textcolor{gray}{\rule[0.25em]{2em}{1.6pt}\ }$;
Statistic $Z_{\hat{\mathcal{L}}}$, $N=50$: $\textcolor{magenta}{\rule[0.25em]{0.4em}{1.6pt} \ \rule[0.25em]{0.4em}{1.6pt}\ \rule[0.25em]{0.4em}{1.6pt} }$;
Statistic $Z_{\hat{\mathcal{L}}}$, $N=100$: $\textcolor{blue}{\rule[0.25em]{1em}{1.6pt} \ \rule[0.25em]{1em}{1.6pt}\ }$; Statistic $Z_{\hat{\mathcal{L}}}$, $N=1000$: $\textcolor{red}{\rule[0.25em]{2em}{1.6pt} }$; 
Statistic $S_{\hat{\mathcal{L}}}$, $N=100$: $\textcolor{black}{\rule[0.25em]{0.6em}{1.7pt} \ \mathbf{\cdot} \ \rule[0.25em]{0.6em}{1.7pt} \ }$
  }
  \label{kernel3}
\end{figure}

Figures \ref{kernel1}-\ref{kernel3} provide kernel estimates of densities of the finite-sample distributions of the statistics $Z_{\hat{\mathcal{L}}}$ and $S_{\hat{\mathcal{L}}}$ for different population distributions and sample sizes.\footnote{The number of replications in all simulation experiments is equal to 100,000.}

Figures \ref{kernel1} and \ref{kernel2} provide kernel density functions of the statistics $Z_{\hat{\mathcal{L}}}$ (sample sizes $N=50, 100, 1000$) and $S_{\hat{\mathcal{L}}}$ (sample size $N=100$)\footnote{Qualitatively similar results for other sample sizes $N$ are omitted for brevity and  available on request.} for, respectively, the empirical Theil and Gini inequality measures in the case of samples from the Singh-Maddala distributions $SM(a_0, b_0, c_0)$ with the parameters $a_0=2.8,$ $b_0=100^{-1/2.8},$ $c_0=1.7$ and the corresponding tail index $\zeta=4.76,$ discussed before. 

In the case of the Theil measure in Figure \ref{kernel1}, we observe some non-Gaussianity in the distribution of the statistics $Z_{\hat{\mathcal{L}}}$ and $S_{\hat{\mathcal{L}}}$ in small and moderate samples. In addition, the density of the $t-$statistic $S_{\hat{\mathcal{L}}}$ for Theil index is considerably (left) skewed in comparison to the densities of the statistic $Z_{\hat{\mathcal{L}}}$. 

In the case of the Gini measure in Figure \ref{kernel2}, we can see that the distribution of the statistic $Z_{\hat{\mathcal{L}}}$ is very close to the standard normal even in small samples. In contrast, the distribution of the $t-$statistic $S$ is again skewed towards left.

For Singh-Maddala distributions with heavier tails as in the case of the parameters $(a, c)=(5.8,0.4473111)$ and the corresponding tail index $\zeta=2.59$ in Figure \ref{kernel3},  the finite sample distributions of the statistics $Z_{\hat{\mathcal{L}}}$ and $S_{\hat{\mathcal{L}}}$ for the Gini measure become more skewed (the same is observed for the Theil measure; the results are omitted for brevity and available on request). Skewness is especially pronounced in the case of small samples and the $t-$statistic $S.$ 

Overall, according to Figures \ref{kernel1}-\ref{kernel3}, normal approximation appears to perform better for finite-sample distributions of the statistic $Z_{\hat{\mathcal{L}}}$ as compared to those of the full-sample $t-$statistic $S_{\hat{\mathcal{L}}}$ used in asymptotic tests and inference. We further note that the group estimators $\hat{\mathcal L}_j-\mathcal{L}_0$ used in $t-$statistic robust inference approaches are just scaled versions of the statistics $Z_{\hat{\mathcal{L}}}$ calculated using observations in the groups considered. Therefore, the above comparisons are expected to translate into better finite-sample performance of $t-$statistic inference approaches 
as compared to the asymptotic ones, provided that the number of observations in each of the groups in $t-$statistic approaches in  sufficiently large, e.g., greater than 100 (this is usually the case in empirical applications with the number of groups $q=2, 4, 8$). For better size control, the number of groups, $q$, should be chosen to be smaller if the total sample size $N$ is not very large.

\subsubsection{Inference in one-sample problem: Finite-sample properties}
Table \ref{tab_1} provides the results on the empirical size of asymptotic and the  $t-$statistic robust tests on the Theil and Gini measures for sample sizes $N=200, 500, 1000$ and Singh-Maddala distributions $SM(a, b_0, c)$ with the parameters $(a, c)=(2.5,2.502199), (3.2,1.2320215), (5.8,0.4996163)$ corresponding to the tail indices $\zeta=6.6, 3.94, 2.9$ in the case of the Theil index and the parameters $(a, c)=(2.5,2.640350), (3.2,1.1866026)$, and $(5.8,0.4473111)$ corresponding to $\zeta=6.26, 3.8, 2.9$ in the case of the Gini index. 


In accordance with the above discussion of finite-sample distributions of statistics $Z$ and $S$ like those in Figures \ref{kernel1}-\ref{kernel3}, the results in the table show that the finite-sample size of both the asymptotic and $t-$statistic robust tests becomes more distorted if the tail index decreases and thus the degree of heavy-tailedness in observations becomes more pronounced. Importantly, however, size distortions for the Gini measure are not so large as for the Theil measure. In the case of the number of groups $q=4$ or $q=8,$ the finite-sample size properties of robust tests based on $t-$statistics in group estimates are usually better than those of the tests based on asymptotic normality of the (full-sample) $t-$statistics for the measures. Further, the finite sample size properties of the robust $t-$statistic-based tests (with $q=4$) appear to be better than those of the asymptotic tests in the cases where each of the groups contains more than 100 observations, in accordance with the discussion of Figures \ref{kernel1}-\ref{kernel3}.
\begin{table}[!h]
\centering
\small
\caption{Empirical size: Identical distributions with $\zeta_I=\zeta_Y=\zeta$ and sample sizes $N_1=N_2=N$\label{tab_1}}
\begin{tabularx}{0.78\textwidth}{cccc|cccc} \toprule
Theil& $\zeta=6.26$ & $\zeta=3.94$ & $\zeta=2.9$ & Gini & $\zeta=6.6$ & $\zeta=3.8$ & $\zeta=2.59$\\ \hline
$N=200$\\
asymptotic &8.2&14.5&25.5&asymptotic&6.2&7.5&13.0\\
$q=$4&6.9&10.6&18.0&$q=4$&5.2&5.2&7.7\\
$q=$8&11.0&17.8&28.7&$q=8$&5.2&6.0&11.3\\
$q=$12&15.9&24.9&37.3&$q=12$&5.5&6.6&14.2\\
$q=$16&21.3&33.1&45.9&$q=16$&5.5&6.9&16.9\\
$N=500$\\
asymptotic &6.9&11.9&20.2&asymptotic&5.7&6.5&10.8\\
$q=$4&5.8&8.2&13.5&$q=4$&4.8&5.1&7.1\\
$q=$8&8.3&12.9&20.6&$q=8$&5.3&5.9&9.5\\
$q=$12&10.0&16.2&25.6&$q=12$&5.1&6.3&11.5\\
$q=$16&12.7&20.0&30.0&$q=16$&5.1&6.4&13.0\\
$N=1000$\\
asymptotic &6.0&9.6&17.0&asymptotic&5.2&5.7&8.6\\
$q=$4&5.3&6.5&10.5&$q=4$&4.8&4.9&5.8\\
$q=$8&6.1&9.3&16.3&$q=8$&4.9&5.1&7.4\\
$q=$12&7.4&11.8&19.5&$q=12$&4.9&5.2&8.5\\
$q=$16&8.7&14.2&22.6&$q=16$&5.0&5.4&10.1\\
  \bottomrule
\end{tabularx}
\end{table}

\subsection{Two-sample problem: Testing equality of two inequality measures and inference on their difference} \label{twosample}
\subsubsection{Inference in two-sample problem: Finite-sample distributions}
In this section, we focus on comparisons of the finite-sample performance of the two-sample $t-$statistic robust inference approaches discussed in Section \ref{tstat} with permutation and bootstrap tests proposed by \cite{Dufour}.

In the numerical analysis in this section, $\mathcal{L}^{I}=\mathcal{L}(F_1)$ and $\mathcal{L}^Y=\mathcal{L}(F_2)$ denote the true values of the inequality measure $\mathcal{L}$ of interest (e.g.,  the Theil or Gini inequality index, as in the previous section) in two populations with cdf's $F_1$ and $F_2$ considered. By $\hat{\mathcal{L}}^I=\hat{\mathcal{L}}^I(I_1, ..., I_{N_1})$ and $\hat{\mathcal{L}}^Y=\hat{\mathcal{L}}^Y(Y_1, ..., Y_{N_2})$ we denote the full-sample estimators of the measure $\mathcal{L}$ (the full-sample empirical inequality measures) calculated using samples of observations $I_1, ..., I_{N_1}$ and $Y_1, ..., Y_{N_2}$
from the two populations. Further, as in Section \ref{tstat}, by $\hat{\mathcal{L}}_1^I, ..., \hat{\mathcal{L}}_{q_1}^I$ and $\hat{\mathcal{L}}_1^Y, ..., \hat{\mathcal{L}}_{q_2}^Y$ we denote the group estimators of $\mathcal{L}$ (group empirical inequality measures) in the two samples dealt with in applications of $t-$statistic inference approaches.

Asymptotic approaches to testing the hypothesis $H_0: L_1-L_2=d_0$ (e.g., with $d_0=0,$ testing the hypothesis $H_0: L_1=L_2$ of equality of the values $L_1, L_2$ of the measure $\mathcal{L}$ in the two populations considered) against the two-sided alternative $H_a: L_1-L_2\neq d_0$ (resp., with $d_0=0,$ against the two-sample alternative $H_a: L_1\neq L_2$) are based on the normal approximation to the sample distribution of the difference $\hat{\mathcal{L}}^I-\hat{\mathcal{L}}^Y$ between the full-sample estimators $\hat{\mathcal{L}}^I$ and $\hat{\mathcal{L}}^Y$ (full-sample empirical inequality measures). More precisely, the asymptotic approaches are based on the standard normal approximation to the sample distribution of the two-sample $t-$statistic  $S_{\hat{\mathcal{L}}_1-\hat{\mathcal{L}}_2}=(\hat{\mathcal{L}}_1-\hat{\mathcal{L}}_2-d_0)/s.e._{\hat{\mathcal{L}}_1-\hat{\mathcal{L}}_2}$ 
calculated using the estimators $\hat{\mathcal{L}}_1$ and $\hat{\mathcal{L}}_2,$ where $s.e._{\hat{\mathcal{L}}_1-\hat{\mathcal{L}}_2}$ denotes the usual consistent standard error of the difference $\hat{\mathcal{L}}_1-\hat{\mathcal{L}}_2$ (see the formulas for the empirical inequality measures considered and the standard errors in \citeauthor{CF}, \citeyear{CF}, \citeauthor{DF}, \citeyear{DF}, \citeauthor{Dufour}, \citeyear{Dufour}). 

Similar to the previous section, validity of two-sample $t-$statistic robust inference approaches based on (\ref{Gen2}) requires weak convergence of group estimators $\hat{\mathcal{L}}_j^I,$ $j=1, ..., q_1,$ $\hat{\mathcal{L}}_k^Y,$ $k=1, ..., q_2,$ of the inequality measure $\mathcal{L}$ in the two samples considered to possibly heterogeneous Gaussian distributions (or scale mixtures of Gaussian distributions). As discussed in the introduction and the previous section, asymptotic normality of group estimators $\hat{\mathcal{L}}_j^I,$ $\hat{\mathcal{L}}_k^Y,$ hold under the same conditions as in the case of the full-sample estimators $\hat{\mathcal{L}}^I$ and $\hat{\mathcal{L}}^Y.$ We refer to the previous section for the assessment of finite-sample distributions of the full-sample empirical inequality measures and their closeness to normality.

On the other hand, with $q_1=q_2=q,$ validity of the (two-sample) $t-$statistic robust inference approaches based on (\ref{GenDiff}) - that is, the one-sample $t-$statistic $\tilde{\tilde{t}}_{\mathcal{L}}$ in the $q$ differences ${\widehat{\mathcal{L}}}^{I}_j-{\widehat{\mathcal{L}}}^{Y}_j,$ $j = 1,...,q,$ of the group empirical inequality measures $\hat{\mathcal{L}}_j^I,$ $\hat{\mathcal{L}}_j^Y$ (without any Studentization/normalization of the differences between the group estimators by their standard errors in contrast to the $t-$statistics $S_{\hat{\mathcal{L}}_1-\hat{\mathcal{L}}_2}$ calculated using the full-sample estimators) requires weak convergence of the differences ${\widehat{\mathcal{L}}}^{I}_j-{\widehat{\mathcal{L}}}^{Y}_j,$ $j = 1,...,q,$ of the group estimators to possibly heterogeneous Gaussian distributions (or scale mixtures of Gaussian distributions). Further, asymptotic normality of the differences $\hat{\mathcal{L}}_j^I-\hat{\mathcal{L}}_j^Y$ between the group estimators  holds under the same conditions as in the case of the difference $\hat{\mathcal{L}}^I-\hat{\mathcal{L}}^Y$ between the full-sample estimators.  

We, therefore, begin the analysis with an assessment of finite-sample distributions of the difference $\hat{\mathcal{L}}^I-\hat{\mathcal{L}}^Y$ between the (full-sample) empirical inequality measures $\hat{\mathcal{L}}^I,$ $\hat{\mathcal{L}}^Y$ and of the (full-sample) $t-$statistics $S_{\hat{\mathcal{L}}^I-\hat{\mathcal{L}}^Y}$ calculated using them. We, in particular, focus on the assessment of closeness of the above finite-sample distributions to Gaussian ones.

We focus on comparisons of finite-sample distributions of the $t-$statistic 
$S_{\hat{\mathcal{L}}^I-\hat{\mathcal{L}}^Y}$ with those of the difference 
$\hat{\mathcal{L}}^I-\hat{\mathcal{L}}^Y$ normalized by its true standard deviation, that is, of the statistic $Z_{\hat{\mathcal{L}}^I-\hat{\mathcal{L}}^Y}=(\hat{\mathcal{L}}^I-\hat{\mathcal{L}}^Y)/\sigma_{\hat{\mathcal{L}}^I-\hat{\mathcal{L}}^Y},$ where $\sigma^2_{\hat{\mathcal{L}}^I-\hat{\mathcal{L}}^Y}=Var(\hat{\mathcal{L}}^I-\hat{\mathcal{L}}^Y).$

\begin{figure}[!h]
  \centering
  \includegraphics[width=0.7\linewidth]{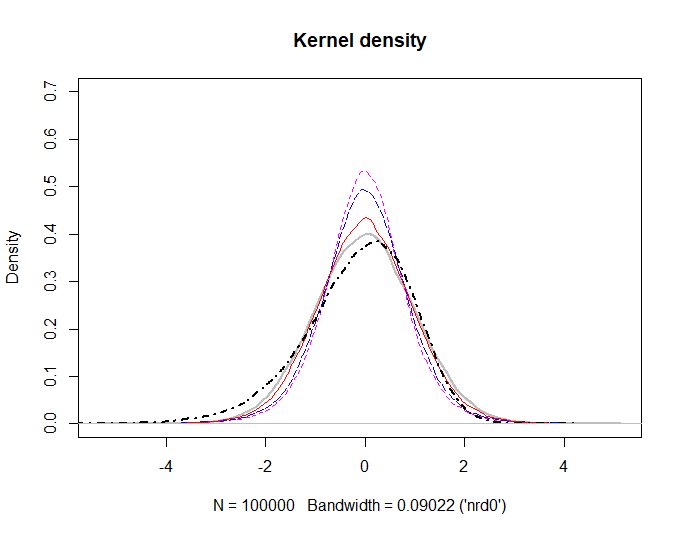}
  \caption{Kernel density functions for the statistics $Z_{\hat{\mathcal{L}}^I-\hat{\mathcal{L}}^Y}$ and $S_{\hat{\mathcal{L}}^I-\hat{\mathcal{L}}^Y}$ for the difference between Theil indices: Singh-Maddala distributions $SM(a_0, b_0, c_0)$ with $(a_0, c_0)=(2.8,1.7)$ and $\zeta=4.76.$ 
  Gaussian density:$\textcolor{gray}{\rule[0.25em]{2em}{1.6pt}\ }$;
Statistic $Z_{\hat{\mathcal{L}}^I-\hat{\mathcal{L}}^Y}$, $N=50$: $\textcolor{magenta}{\rule[0.25em]{0.4em}{1.6pt} \ \rule[0.25em]{0.4em}{1.6pt}\ \rule[0.25em]{0.4em}{1.6pt} }$;
Statistic $Z_{\hat{\mathcal{L}}^I-\hat{\mathcal{L}}^Y}$, $N=100$: $\textcolor{blue}{\rule[0.25em]{1em}{1.6pt} \ \rule[0.25em]{1em}{1.6pt}\ }$; Statistic $Z_{\hat{\mathcal{L}}^I-\hat{\mathcal{L}}^Y}$, $N=1000$: $\textcolor{red}{\rule[0.25em]{2em}{1.6pt} }$; 
Statistic $S_{\hat{\mathcal{L}}^I-\hat{\mathcal{L}}^Y}$, $N=100$: $\textcolor{black}{\rule[0.25em]{0.6em}{1.7pt} \ \mathbf{\cdot} \ \rule[0.25em]{0.6em}{1.7pt} \ }$
  }
  \label{kernel4}
\end{figure}

\begin{figure}[!h]
  \centering
  \includegraphics[width=0.7\linewidth]{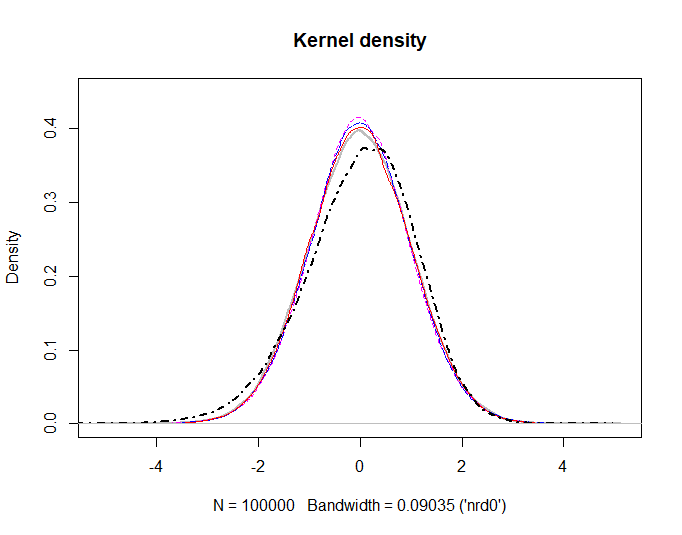}
  \caption{Kernel density functions for the statistics $Z_{\hat{\mathcal{L}}^I-\hat{\mathcal{L}}^Y}$ and $S_{\hat{\mathcal{L}}^I-\hat{\mathcal{L}}^Y}$ for the difference between Gini indices: Singh-Maddala distributions $SM(a_0, b_0, c_0)$ with $(a_0, c_0)=(2.8,1.7)$ and $\zeta=4.76.$. 
  Gaussian density:$\textcolor{gray}{\rule[0.25em]{2em}{1.6pt}\ }$;
Statistic $Z_{\hat{\mathcal{L}}^I-\hat{\mathcal{L}}^Y}$, $N=50$: $\textcolor{magenta}{\rule[0.25em]{0.4em}{1.6pt} \ \rule[0.25em]{0.4em}{1.6pt}\ \rule[0.25em]{0.4em}{1.6pt} }$;
Statistic $Z_{\hat{\mathcal{L}}^I-\hat{\mathcal{L}}^Y}$, $N=100$: $\textcolor{blue}{\rule[0.25em]{1em}{1.6pt} \ \rule[0.25em]{1em}{1.6pt}\ }$;  Statistic $Z_{\hat{\mathcal{L}}^I-\hat{\mathcal{L}}^Y}$, $N=1000$: $\textcolor{red}{\rule[0.25em]{2em}{1.6pt} }$, 
Statistic $S_{\hat{\mathcal{L}}^I-\hat{\mathcal{L}}^Y}$, $N=100$: $\textcolor{black}{\rule[0.25em]{0.6em}{1.7pt} \ \mathbf{\cdot} \ \rule[0.25em]{0.6em}{1.7pt} \ }$
  }
  \label{kernel5}
\end{figure}

\begin{figure}[!h]
  \centering
  \includegraphics[width=0.7\linewidth]{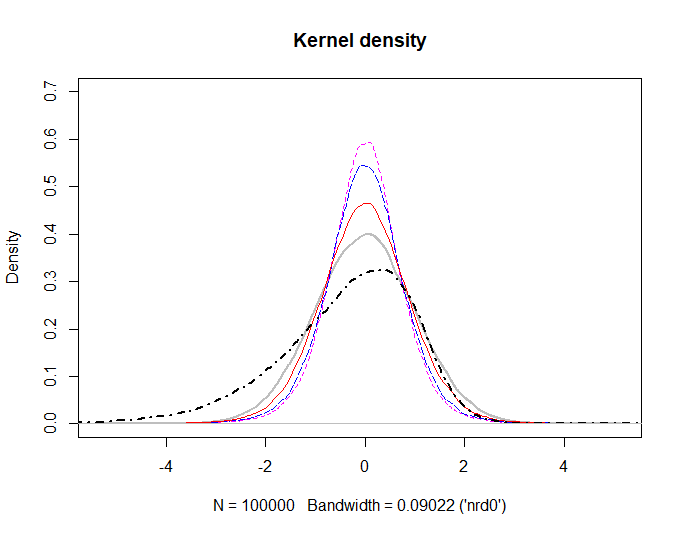}
  \caption{Kernel density functions for the statistics $Z_{\hat{\mathcal{L}}^I-\hat{\mathcal{L}}^Y}$ and $S_{\hat{\mathcal{L}}^I-\hat{\mathcal{L}}^Y}$ for the difference between Gini indices: Singh-Maddala distributions $SM(a, b_0, c)$ with $(a, c)=(5.8,0.4473111)$ and  $\zeta=2.59.$ 
  Gaussian density:$\textcolor{gray}{\rule[0.25em]{2em}{1.6pt}\ }$;
Statistic $Z_{\hat{\mathcal{L}}^I-\hat{\mathcal{L}}^Y}$, $N=50$: $\textcolor{magenta}{\rule[0.25em]{0.4em}{1.6pt} \ \rule[0.25em]{0.4em}{1.6pt}\ \rule[0.25em]{0.4em}{1.6pt} }$;
Statistic $Z_{\hat{\mathcal{L}}^I-\hat{\mathcal{L}}^Y}$, $N=100$: $\textcolor{blue}{\rule[0.25em]{1em}{1.6pt} \ \rule[0.25em]{1em}{1.6pt}\ }$;  Statistic $Z_{\hat{\mathcal{L}}^I-\hat{\mathcal{L}}^Y}$, $N=1000$: $\textcolor{red}{\rule[0.25em]{2em}{1.6pt} }$; Statistic 
$S_{\hat{\mathcal{L}}^I-\hat{\mathcal{L}}^Y},$ $N=100$: $\textcolor{black}{\rule[0.25em]{0.6em}{1.7pt} \ \mathbf{\cdot} \ \rule[0.25em]{0.6em}{1.7pt} \ }$
  }
  \label{kernel6}
\end{figure}

In Figures \ref{kernel4}-\ref{kernel6}, we present kernel estimates of the finite-sample densities of the statistics $Z_{\hat{\mathcal{L}}^I-\hat{\mathcal{L}}^Y}$ and $S_{\hat{\mathcal{L}}^I-\hat{\mathcal{L}}^Y}$ for the difference between the empirical inequality measures in two samples from populations with the same Singh-Maddala distribution.

Figures \ref{kernel4}-\ref{kernel5} provide kernel density functions, for sample sizes $N_1=N_2=N,$ of the statistics $Z_{\hat{\mathcal{L}}^I-\hat{\mathcal{L}}^Y}$ ($N=50, 100, 1000$) and $S_{\hat{\mathcal{L}}^I-\hat{\mathcal{L}}^Y}$ (sample sizes $N=100$)\footnote{Qualitatively similar results for other sample sizes $N$ are omitted for brevity and  available on request.} for the difference between, respectively, the Theil and Gini empirical inequality measures in two samples from the Singh-Maddala distribution $SM(a_0, b_0, c_0)$ with the parameters $a_0=2.8,$ $b_0=100^{-1/2.8},$ $c_0=1.7$ and the corresponding tail index $\zeta=4.76.$ Figure \ref{kernel6} provides the above kernel density functions for the Gini index in the case of two samples from a more heavy-tailed Singh-Maddala distribiion $SM(a, b_0, c)$ with $(a, c)=(5.8,0.4473111)$ and the tail index $\zeta=2.59.$




According to Figures \ref{kernel4}-\ref{kernel6}, the finite-sample distributions of the statistic $Z_{\hat{\mathcal{L}}^I-\hat{\mathcal{L}}^Y}$ and thus of the difference $\hat{\mathcal{L}}^I-\hat{\mathcal{L}}^Y$ between the estimators of the Theil and Gini measures is approximately symmetric even in rather small samples and also under pronounced heavy-tailedness, with good performance of Gaussian approximations, e.g., as compared to finite-sample distribution of the (full-sample) $t-$statistic $S_{\hat{\mathcal{L}}^I-\hat{\mathcal{L}}^Y}.$ This also holds in the case when the sample sizes are not very different. \footnote{If the sample sizes of two groups are very different, then different partition, $q_1,q_2$ should be used in applications of the $t$-statistic inference approaches.}




\subsubsection{Inference in two-sample problem: Finite-sample size properties}

Tables \ref{tab_2_1}-\ref{tab_2_6} provide the results on the finite-sample size properties of the asymptotic, permutation, bootstrap and $t-$statistic robust tests on equality of Theil and Gini measures. As before, we consider two samples $I_1, ..., I_{N_1}$ and $Y_1, ..., Y_{N_2}$ from, respectively, Singh-Maddala distributions $SM(a_I, b_0, c_I)$ and $SM(a_Y, b_0, c_Y),$  with $b_0=100^{-1/2.8}$ and the tail indices $\zeta_I=a_Ic_I,$ $\zeta_Y=a_Yc_Y.$ In simulations, we consider the following settings with identical/different sample sizes $N_1,$ $N_2;$ distributions $S(a_I, b_0, c_I)$ and $S(a_Y, b_0, c_Y)$ in the samples and the number $q_1, q_2$ of groups used in $t-$statistic robust tests. 

Tables \ref{tab_2_1}-\ref{tab_2_5} provide the results on finite-sample size properties of asymptotic, permutation, bootstrap and $t-$statistic robust tests based on  (\ref{Gen2}) and (\ref{GenDiff}) with the equal number of groups $q_1=q_2=q.$

\begin{enumerate}[label=(\roman*)]
\item \label{settings} Identical distributions and sample sizes $N_1=N_2=N=200$ (Table \ref{tab_2_1}). 

The values of the parameters $(a, c)$ as in \cite{Dufour} and Table \ref{tab_2_1}:
$(a_I, c_I)=(a_Y, c_Y)=(2.5,2.502199), (3.2,1.2320215), (5.8,0.4996163)$ and $\zeta_I=\zeta_Y=\zeta=6.6, 3.94, 2.9$ (Theil index);  $(a_I, c_I)=(a_Y, c_Y)=(2.5,2.640350), (3.2,1.1866026), (5.8,0.4473111)$ and $\zeta_I=\zeta_Y=\zeta=6.6, 3.8, 2.59$ (Gini index);

Singh-Maddala distributions with tail indices $\zeta$ similar to the empirical results in the literature with $\zeta\in (1.5, 3)$ for income and $\zeta\approx 1.5$ for wealth (see Section \ref{num} and references therein):

$(a_I, c_I)=(a_Y, c_Y)=(2, 1.1)$ and $\zeta_I=\zeta_Y=\zeta=2.2;$  $(a_I, c_I)=(a_Y, c_Y)=(2, 0.7)$ and $\zeta_I=\zeta_Y=\zeta=1.4$ (Theil and Gini measures).

\end{enumerate}

The results in Table \ref{tab_2_1} indicate that the size of all the tests, except the asymptotic ones, never exceeds the nominal 5\% level. In addition, in a number of cases, it is quite close to the nominal level for the permutation, bootstrap and the robust $t-$statistic tests. 

\begin{enumerate}[label=(\roman*)] \setcounter{enumi}{1}
\item \label{settings1} Identical sample sizes $N_1=N_2=N=200$ and different distributions (Table \ref{tab_2_2}). 

$(a_I, c_I)=(2.8, 1.7)$ and $\zeta_I=4.76;$

$(a_Y, c_Y)$=$(2.5, 2.502199), (3.2, 1.2320215), (5.8, 0.4996163);$ $\zeta_Y=2.9, 3.94, 6.6$ (Theil index);

$(a_Y, c_Y)$=$(2.5,2.640350), (3.2, 1.1866026), (5.8, 0.4473111);$ $\zeta_Y=2.9, 3.8, 6.26$ (Gini index).

\end{enumerate}
According to Table \ref{tab_2_2}, the empirical sizes of the robust 
two-sample $t-$statistic tests based on (\ref{Gen2}) with $q=4, 8$ in the case of more heavy-tailed distributions and on (\ref{GenDiff}) with $q=4, 8, 12, 16$ in the case of less heavy-tailed distributions are comparable and in some cases are better than those of the permutation and bootstrap tests. Comparing the robust tests based on the two-sample $t-$statistic (\ref{Gen2}) in group estimators and those based on the one-sample $t-$statistic (\ref{GenDiff}) in the differences of the group estimators, overall, the former tests with the same number of groups $q_1=q_2=q$ appear to have less over-rejections as compared to the latter ones.

\begin{enumerate}[label=(\roman*)] \setcounter{enumi}{2}
\item \label{settings2} Different sample sizes $N_1, N_2$ and identical distributions (Tables \ref{tab_2_3} and \ref{tab_2_4}).

$N_1=200,$ $N_2=50,200,500,1000,5000;$ 

$(a_I, c_I)=(a_Y, c_Y)=(5.8,0.4996163)$ and $\zeta_I=\zeta_Y=2.9$ in the case of the Theil index;  $(a_I, c_I)=(a_Y, c_Y)=(5.8,0.4473111)$ and $\zeta_I=\zeta_Y=2.59$ in the case of the Gini index (Table \ref{tab_2_3});

$(a_I, c_I)=(a_Y, c_Y)=(2, 0.7)$ and $\zeta_I=\zeta_Y=1.4$ (Theil and Gini measures, Table \ref{tab_2_4}).

\end{enumerate}


The finite-sample size of all the tests, except the asymptotic ones, appears to be good in all parameter settings: e.g., essentially no over-rejections are observed for $t-$statistic inference approaches, including the settings with more pronounced heavy-tailedness and infinite variances in Table \ref{tab_2_4}. Further, the finite sample properties of the robust $t-$statistic approaches are comparable or similar and in some cases are better than those of the bootstrap and permutation approaches. Importantly, asymptotic normality of sample Theil and Gini measures is lost under infinite variances, as is the case for tail indices $\zeta_I=\zeta_Y=1.4$ in Table \ref{tab_2_4} (see \cite{Taleb} and the discussion in Appendix B). However, according to the results in the table, the $t-$statistic approaches have good finite sample size properties even in such heavy-tailed settings. This is due to robustness of $t-$statistic approaches to heavy-tailedness as they may be used under convergence of group estimators of parameters in consideration to scale mixtures of normals.

\begin{enumerate}[label=(\roman*)] \setcounter{enumi}{2}
\item \label{settings3} Different sample sizes $N_1, N_2$ and different distributions (Table \ref{tab_2_5}).

$N_1=200,$ $N_2=50,200,500,1000,5000;$ 

$(a_I, c_I)=(2.8, 1.7)$ and $\zeta_I=4.76;$

$(a_Y, c_Y)=(5.8,0.4996163)$ and $\zeta_Y=2.9$ (Theil index);  $(a_Y, c_Y)=(5.8,0.4473111)$ and $\zeta_Y=2.59$ (Gini index).



\end{enumerate}

In Table \ref{tab_2_5}, one can observe better size properties for two-sample $t-$statistic inference approaches based on $\tilde{t}_{\mathcal{L}}$ in (\ref{Gen2}) (with $q_1=q_2=4,8,12$ for all sample sizes and also $q=16$ for large sample sizes) and those based on $\tilde{\tilde{t}}_{\mathcal{L}}$ in (\ref{GenDiff}) (with $q_1=q_2=4,8$ for all sample sizes) in comparison to permutation and bootstrap tests.

\begin{table}[!h]
\centering
\small
\caption{Empirical size, identical distributions with $\zeta_I=\zeta_Y=\zeta$ and sample sizes $N_1=N_2=200$ \label{tab_2_1}}
\begin{tabularx}{0.95\textwidth}{cccccc|cccccc} \toprule
Theil \textbackslash $\zeta$&6.26&3.94&2.9&2.2&1.4&Gini\textbackslash $\zeta$ &6.6&3.8&2.59&2.2&1.4\\\hline
asymptotic &5.4&5.5&5.3&7.6&22.1&asy&5.7&6.1&7.0&8.4&18.8\\
$\tilde{t}_{\mathcal{L}}(q=4)$&2.0&1.7&1.5&1.5&1.3&$\tilde{t}_{\mathcal{L}}(q=4)$&2.1&1.9&1.9&2.0&2.0\\
$\tilde{t}_{\mathcal{L}}(q=8)$&3.2&2.7&2.2&2.4&2.4&$\tilde{t}_{\mathcal{L}}(q=8)$&3.5&3.4&3.0&3.2&3.4\\
$\tilde{t}_{\mathcal{L}}(q=12)$&3.6&3.2&2.7&2.9&3.0&$\tilde{t}_{\mathcal{L}}(q=12)$&3.8&3.7&3.4&3.7&3.7\\
$\tilde{t}_{\mathcal{L}}(q=16)$&4.0&3.7&3.1&3.4&3.5&$\tilde{t}_{\mathcal{L}}(q=16)$&4.4&4.2&4.1&4.0&4.1\\
$\tilde{\tilde{t}}_{\mathcal{L}}(q=4)$&4.6&4.4&3.5&3.5&2.9&$\tilde{\tilde{t}}_{\mathcal{L}}(q=4)$&5.0&4.9&4.4&4.6&4.6\\
$\tilde{\tilde{t}}_{\mathcal{L}}(q=8)$&4.7&4.1&3.3&3.7&3.7&$\tilde{\tilde{t}}_{\mathcal{L}}(q=8)$&4.9&4.8&4.5&4.9&4.9\\
$\tilde{\tilde{t}}_{\mathcal{L}}(q=12)$&4.9&4.3&3.5&3.8&4.1&$\tilde{\tilde{t}}_{\mathcal{L}}(q=12)$&5.1&4.9&4.6&4.8&4.9\\
$\tilde{\tilde{t}}_{\mathcal{L}}(q=16)$&5.0&4.5&3.8&4.1&4.3&$\tilde{\tilde{t}}_{\mathcal{L}}(q=16)$&5.3&5.1&4.7&4.8&5.0\\
permutation&4.8&4.7&4.9&4.7&4.7&permutation&4.4&4.5&4.8&4.8&4.5\\
bootstrap&4.5&4.2&3.4&3.3&2.8&bootstrap&4.8&4.4&4.1&3.9&3.9\\
  \bottomrule
\end{tabularx}
\end{table}

\begin{table}[!h]
\centering
\small
\caption{Empirical size, identical sample sizes $N_1=N_2=200$ and different distributions, $\zeta_I=4.76$ \label{tab_2_2}}
\begin{tabularx}{0.63\textwidth}{cccc|cccc} \toprule
Theil\textbackslash $\zeta_Y$ &6.26&3.94&2.9&Gini \textbackslash $\zeta_Y$ &6.6&3.8&2.59\\\hline
asy&5.1&5.4&12.3&asy&5.2&5.5&8.0\\
$\tilde{t}_{\mathcal{L}}(q=4)$&1.9&1.5&4.0&$\tilde{t}_{\mathcal{L}}(q=4)$&1.9&1.7&2.7\\
$\tilde{t}_{\mathcal{L}}(q=8)$&3.1&2.9&8.7&$\tilde{t}_{\mathcal{L}}(q=8)$&3.2&3.1&5.1\\
$\tilde{t}_{\mathcal{L}}(q=12)$&3.6&3.4&11.4&$\tilde{t}_{\mathcal{L}}(q=12)$&3.8&3.6&6.6\\
$\tilde{t}_{\mathcal{L}}(q=16)$&4.0&3.9&13.6&$\tilde{t}_{\mathcal{L}}(q=16)$&3.9&3.9&7.8\\
$\tilde{\tilde{t}}_{\mathcal{L}}(q=4)$&4.8&4.3&7.2&$\tilde{\tilde{t}}_{\mathcal{L}}(q=4)$&5.1&4.7&5.7\\
$\tilde{\tilde{t}}_{\mathcal{L}}(q=8)$&5.0&4.5&10.6&$\tilde{\tilde{t}}_{\mathcal{L}}(q=8)$&5.1&4.8&6.6\\
$\tilde{\tilde{t}}_{\mathcal{L}}(q=12)$&4.8&4.6&12.6&$\tilde{\tilde{t}}_{\mathcal{L}}(q=12)$&5.0&4.8&8.1\\
$\tilde{\tilde{t}}_{\mathcal{L}}(q=16)$&5.1&4.8&14.5&$\tilde{\tilde{t}}_{\mathcal{L}}(q=16)$&5.0&4.8&8.8\\
permutation&4.8&4.8&11.0&permutation&4.4&4.6&6.2\\
bootstrap&4.8&4.7&10.6&bootstrap&4.6&4.4&6.3\\
  \bottomrule
\end{tabularx}
\end{table}


\begin{table}[!h]
\centering
\small
\caption{Empirical size, identical distributions with $\zeta_I=\zeta_Y=\zeta$ and different sample sizes, $N_1=200$ \label{tab_2_3}}
\begin{tabularx}{0.98\textwidth}{cccccc|cccccc} \toprule
Theil, $\zeta=2.9$ \textbackslash $N_2$& 50 & 200 & 500 & 1000 & 5000 & Gini, $\zeta=2.59$\textbackslash $N_2$ &50&200&500&1000&5000\\\hline
asymptotic &8.7&5.3&4.7&4.6&4.5&asy&12.3&6.6&5.7&5.4&5.2\\
$\tilde{t}_{\mathcal{L}}(q=4)$&1.2&1.2&1.6&1.5&1.6&$\tilde{t}_{\mathcal{L}}(q=4)$&1.5&1.5&1.8&1.8&1.7\\
$\tilde{t}_{\mathcal{L}}(q=8)$&2.2&2.0&2.4&2.4&2.5&$\tilde{t}_{\mathcal{L}}(q=8)$&3.1&2.7&3.2&3.1&3.4\\
$\tilde{t}_{\mathcal{L}}(q=12)$&2.7&2.4&2.9&2.8&3.0&$\tilde{t}_{\mathcal{L}}(q=12)$&3.6&3.4&3.5&3.6&4.0\\
$\tilde{t}_{\mathcal{L}}(q=16)$&3.5&2.8&3.2&3.3&3.4&$\tilde{t}_{\mathcal{L}}(q=16)$&4.3&3.7&3.6&3.9&4.2\\
$\tilde{\tilde{t}}_{\mathcal{L}}(q=4)$&2.9&3.1&3.7&3.7&3.7&$\tilde{\tilde{t}}_{\mathcal{L}}(q=4)$&4.3&4.4&4.8&4.7&4.8\\
$\tilde{\tilde{t}}_{\mathcal{L}}(q=8)$&3.6&3.2&3.9&3.7&4.0&$\tilde{\tilde{t}}_{\mathcal{L}}(q=8)$&4.5&4.4&4.8&4.7&4.6\\
$\tilde{\tilde{t}}_{\mathcal{L}}(q=12)$&3.8&3.7&3.8&3.7&4.4&$\tilde{\tilde{t}}_{\mathcal{L}}(q=12)$&4.9&4.6&4.7&4.9&5.1\\
$\tilde{\tilde{t}}_{\mathcal{L}}(q=16)$&4.5&3.5&3.8&3.9&4.0&$\tilde{\tilde{t}}_{\mathcal{L}}(q=16)$&5.2&4.6&4.7&4.7&4.8\\
permutation&5.0&4.8&4.9&4.9&5.0&permutation&5.1&4.9&5.0&4.9&5.2\\
bootstrap&4.0&4.2&4.1&4.5&4.7&bootstrap&4.6&4.8&4.8&5.1&5.3\\
  \bottomrule
\end{tabularx}
\end{table}

\begin{table}[!h]
\centering
\small
\caption{Empirical size, identical distributions with $\zeta_I=\zeta_Y=1.4$ and different sample sizes, $N_1=200$ \label{tab_2_4}}
\begin{tabularx}{0.92\textwidth}{cccccc|cccccc} \toprule
Theil\textbackslash $N_2$ &50&200&500&1000&5000&Gini \textbackslash $N_2$ &50&200&500&1000&5000\\\hline
asymptotic &31.5&21.9&16.7&13.8&8.7&asymptotic &28.9&18.3&14.0&12.3&8.6\\
$\tilde{t}_{\mathcal{L}}(q=4)$&1.0&1.1&1.3&1.1&1.1&$\tilde{t}_{\mathcal{L}}(q=4)$&1.7&1.7&1.9&1.6&1.6\\
$\tilde{t}_{\mathcal{L}}(q=8)$&2.5&2.2&2.3&2.1&2.0&$\tilde{t}_{\mathcal{L}}(q=8)$&3.3&3.0&3.1&3.2&2.7\\
$\tilde{t}_{\mathcal{L}}(q=12)$&3.5&3.0&3.1&2.9&2.7&$\tilde{t}_{\mathcal{L}}(q=12)$&3.8&3.5&3.8&3.7&3.5\\
$\tilde{t}_{\mathcal{L}}(q=16)$&4.0&3.3&3.3&3.5&3.0&$\tilde{t}_{\mathcal{L}}(q=16)$&4.3&3.8&4.1&4.0&4.0\\
$\tilde{\tilde{t}}_{\mathcal{L}}(q=4)$&3.0&3.0&3.1&2.8&2.7&$\tilde{\tilde{t}}_{\mathcal{L}}(q=4)$&4.7&4.5&4.5&4.3&3.9\\
$\tilde{\tilde{t}}_{\mathcal{L}}(q=8)$&4.0&3.5&3.6&3.5&3.2&$\tilde{\tilde{t}}_{\mathcal{L}}(q=8)$&5.2&4.8&4.9&4.7&4.3\\
$\tilde{\tilde{t}}_{\mathcal{L}}(q=12)$&4.7&4.1&3.9&3.8&3.7&$\tilde{\tilde{t}}_{\mathcal{L}}(q=12)$&5.1&5.0&4.9&4.9&4.8\\
$\tilde{\tilde{t}}_{\mathcal{L}}(q=16)$&5.2&4.2&4.1&4.2&3.7&$\tilde{\tilde{t}}_{\mathcal{L}}(q=16)$&5.4&4.9&4.7&5.0&4.6\\
permutation&4.9&4.8&5.1&5.1&4.9&permutation&5.1&4.7&4.9&5.0&5.0\\
bootstrap&3.3&3.4&3.6&3.6&3.6&bootstrap&4.5&4.2&4.3&4.4&4.1\\
  \bottomrule
\end{tabularx}
\end{table}




\begin{table}[!h]
\centering
\small
\caption{Empirical size, different distributions and sample sizes, $\zeta_I=4.76,$ $N_1=200$ \label{tab_2_5}}
\begin{tabularx}{0.99\textwidth}{cccccc|cccccc} \toprule
Theil, $\zeta_Y=2.9$ \textbackslash $N_2$&50&200&500&1000&5000&Gini, $\zeta_Y=2.59$\textbackslash $N_2$ &50&200&500&1000&5000\\\hline
asymptotic &14.5&12.3&12.2&11.2&9.0&asymptotic &12.6&8.0&7.5&6.6&6.0\\
$\tilde{t}_{\mathcal{L}}(q=4)$&3.9&4.0&4.8&4.2&4.0&$\tilde{t}_{\mathcal{L}}(q=4)$&2.8&2.7&2.9&2.7&2.6\\
$\tilde{t}_{\mathcal{L}}(q=8)$&8.8&8.7&8.6&8.0&6.7&$\tilde{t}_{\mathcal{L}}(q=8)$&6.0&5.1&5.1&4.8&4.3\\
$\tilde{t}_{\mathcal{L}}(q=12)$&11.4&11.4&11.3&9.7&7.8&$\tilde{t}_{\mathcal{L}}(q=12)$&7.7&6.6&6.6&5.8&5.1\\
$\tilde{t}_{\mathcal{L}}(q=16)$&13.9&13.6&12.8&11.2&8.4&$\tilde{t}_{\mathcal{L}}(q=16)$&9.8&7.8&7.5&6.6&5.5\\
$\tilde{\tilde{t}}_{\mathcal{L}}(q=4)$&6.9&7.2&7.8&7.3&6.6&$\tilde{\tilde{t}}_{\mathcal{L}}(q=4)$&5.7&5.7&6.1&5.7&5.2\\
$\tilde{\tilde{t}}_{\mathcal{L}}(q=8)$&11.1&10.6&10.2&9.6&7.9&$\tilde{\tilde{t}}_{\mathcal{L}}(q=8)$&8.2&6.6&6.8&6.2&5.6\\
$\tilde{\tilde{t}}_{\mathcal{L}}(q=12)$&13.0&12.6&12.2&10.9&8.7&$\tilde{\tilde{t}}_{\mathcal{L}}(q=12)$&9.3&8.1&7.7&6.9&6.2\\
$\tilde{\tilde{t}}_{\mathcal{L}}(q=16)$&14.9&14.5&13.6&12.1&9.0&$\tilde{\tilde{t}}_{\mathcal{L}}(q=16)$&11.0&8.8&8.2&7.4&6.2\\
permutation&12.4&11.0&11.6&10.5&8.8&permutation&8.9&6.2&6.5&6.1&5.6\\
bootstrap&12.2&10.6&11.1&10.4&8.5&bootstrap&9.1&6.3&6.3&6.0&5.7\\
  \bottomrule
\end{tabularx}
\end{table}


Table \ref{tab_2_6} is an analogue of Tables \ref{tab_2_3} and \ref{tab_2_4}  with different numbers $q_1, q_2$ of groups used in two-sample $t-$statistic robust inference approaches based on $\tilde{t}_{\mathcal{L}}$ in (\ref{Gen2}). It provides the results on the empirical size of the tests based on these approaches with asymptotic, bootstrap and permutation tests in the following settings.

\begin{enumerate}[label=(\roman*)] \setcounter{enumi}{2}
\item \label{settings5} Identical distributions and different sample sizes $N_1, N_2$ and the numbers $q_1, q_2$ of groups

$(a_I, c_I)=(a_Y, c_Y)=(2, 1.1),$ $\zeta_I=\zeta_Y=2.2;$ 

$N_1=200,$ $N_2=400, 600, 800.$ 



\end{enumerate}


According to the results in Table \ref{tab_2_6}, in the case of two-sample $t-$statistic inference on equality of/the difference between Theil indices, only the choice of $q_1=q_2=4$ leads to size control for all sample sizes considered. Size distortion of the $t-$statistic approaches in the case of Theil indices is apparently due to skewness in finite-sample distributions of (group) empirical Theil inequality measures implying poor quality of normal approximations to them (see Section \ref{fs}). The solution may be to use different number of groups $q_1, q_2$ for different sample size pairs $N_1, N_2$.  
E.g., according to Table \ref{tab_2_6}, in the case of inference on equality/the difference between the Theil indices, good size properties of two-sample $t-$statistic approaches are observed with $(q_1, q_2)=(8, 8)$ for $N_1=200, N_2=400$, $(q_1, q_2)=(6, 8)$ for $N_1=200, N_2=600$ and $(q_1, q_2)=(6, 12)$ for $N_1=200, N_2=800$. The finite-sample distributions of (group) empirical Gini measures are not so skewed and better approximated by normal ones as compared to the Theil measures (see Section \ref{fs}). In Table \ref{tab_2_6}, one observes good size control for different combinations of $q_1$ and $q_2$ in $t-$statistic robust tests of equality of/the difference between the Gini indices except only the cases $q_1=q_2=12$,  $q_1=q_2=16$ and $q_1=12, q_2=16$ with rather small number of observations in each of the group. To avoid very conservative size properties, the best choices for the number of groups in applications of $t-$statistic robust tests in the case of Gini indices appear to be $(q_1, q_2)=(8, 8)$, $(q_1, q_2)=(9, 12)$ and $(q_1, q_2)=(8, 16)$ for all sample sizes $N_1, N_2$ considered.

\begin{table}[!h]
\centering
\small
\caption{Empirical size, identical distributions with $\zeta_I=\zeta_Y=2.2$ and different sample sizes, $N_1=200$ \label{tab_2_6}}
\begin{tabularx}{0.76\textwidth}{cccc|cccc} \toprule
Theil \textbackslash $N_2$&400&600&800&Gini\textbackslash $N_2$ &400&600&800\\\hline
asymptotic &8.8&9.3&11.4&asymptotic &8.8&8.1&9.1\\
$\tilde{t}_{\mathcal{L}}(q_1=4,q_2=4)$&1.9&2.3&3.2&$\tilde{t}_{\mathcal{L}}(q_1=4,q_2=4)$&2.2&2.1&2.7\\
$\tilde{t}_{\mathcal{L}}(q_1=8,q_2=8)$&4.2&6.3&8.6&$\tilde{t}_{\mathcal{L}}(q_1=8,q_2=8)$&3.9&4.3&5.1\\
$\tilde{t}_{\mathcal{L}}(q_1=12,q_2=12)$&6.8&11.1&15.1&$\tilde{t}_{\mathcal{L}}(q_1=12,q_2=12)$&5.1&5.7&6.8\\
$\tilde{t}_{\mathcal{L}}(q_1=16,q_2=16)$&9.4&15.9&21.6&$\tilde{t}_{\mathcal{L}}(q_1=16,q_2=16)$&5.8&6.6&8.7\\
$\tilde{t}_{\mathcal{L}}(q_1=3,q_2=4)$&0.7&0.7&1.1&$\tilde{t}_{\mathcal{L}}(q_1=3,q_2=4)$&0.8&1.0&1.3\\
$\tilde{t}_{\mathcal{L}}(q_1=6,q_2=8)$&2.5&3.6&5.5&$\tilde{t}_{\mathcal{L}}(q_1=6,q_2=8)$&2.8&3.1&3.8\\
$\tilde{t}_{\mathcal{L}}(q_1=9,q_2=12)$&4.1&6.3&9.4&$\tilde{t}_{\mathcal{L}}(q_1=9,q_2=12)$&3.9&4.1&5.0\\
$\tilde{t}_{\mathcal{L}}(q_1=12,q_2=16)$&6.0&9.5&14.0&$\tilde{t}_{\mathcal{L}}(q_1=12,q_2=16)$&4.8&5.1&6.4\\
$\tilde{t}_{\mathcal{L}}(q_1=2,q_2=4)$&0.1&0.1&0.1&$\tilde{t}_{\mathcal{L}}(q_1=2,q_2=4)$&0.0&0.1&0.1\\
$\tilde{t}_{\mathcal{L}}(q_1=4,q_2=8)$&1.1&1.1&1.9&$\tilde{t}_{\mathcal{L}}(q_1=4,q_2=8)$&1.4&1.5&2.0\\
$\tilde{t}_{\mathcal{L}}(q_1=6,q_2=12)$&2.1&2.9&4.6&$\tilde{t}_{\mathcal{L}}(q_1=6,q_2=12)$&2.6&2.8&3.4\\
$\tilde{t}_{\mathcal{L}}(q_1=8,q_2=168)$&3.1&4.5&6.9&$\tilde{t}_{\mathcal{L}}(q_1=8,q_2=16)$&3.4&3.6&4.5\\
permutation&5.4&4.9&4.9&permutation&5.4&4.7&4.8\\
bootstrap&4.6&3.7&4.0&bootstrap&5.3&4.4&4.4\\
  \bottomrule
\end{tabularx}
\end{table}

Finally, Table \ref{tab_spatial} provides the results for the case of samples with dependent observations, i.e., those with spatially dependent data relevant for studies of income distributions and inequality. Each of the two samples consists of 192 observations with the standard (parameters $\mu=0$ and $\sigma=1$) lognormal distribution located on a rectangular array of unit squares with 16 rows and 12 columns. The observations are generated such that the correlation between the logarithms of two observations is given by $\exp(-\phi d)$ for some $\phi>0,$ where $d$ is the Euclidean distance between the two observations (see Section 3.4 in \cite{IM2} for the use of a similar spatially correlated setting in the analysis of finite sample size properties of one-sample $t-$statistic approaches in inference on the mean of Gaussian observations with spatial dependence). The case $\phi=\infty$ to samples of i.i.d. observations.  

More precisely, the observations in the samples are given by $I_{ij}=\exp(u_{ij}),$ $Y_{ij}=\exp(v_{ij}),$ $i=1, ..., 16,$ $j=1, ..., 12,$ where $u_{ij}$ and $v_{ij}$ are multivariate mean zero unit variance Gaussian with correlation between $u_{ij}$ and $u_{lk}$ and between $v_{ij}$ and $v_{lk}$ equals $\exp(-\phi\sqrt{(i-l)^2+(j-k)^2}).$ 

According to the results in Table \ref{tab_spatial}, the empirical size properties of $t-$statistic tests of equality of Theil and Gini indices in the two samples with spatial dependence are comparable (especially, for the tests based on the two-sample $t-$statistic $\tilde{\tilde t}$ with $q_1=q_2=q=4$ groups and the one-sample $t-$statistic $\tilde t$ in differences with $q=8$) to those of permutation and bootstrap procedures. Furthermore, the finite sample size properties of essentially all robust $t-$statistic tests are better than those of bootstrap and permutation tests under pronounced spatial dependence with $\phi=1.$

\begin{table}[!h]
\centering
\small
\caption{Empirical size, spatial correlation \label{tab_spatial}}
\begin{tabularx}{0.6\textwidth}{cccc|cccc} \toprule
Theil \textbackslash $\phi$&$\infty$&2&1&Gini\textbackslash $\phi$ & $\infty$&2&1\\\hline
asymptotic &7.1&7.6&14.6&asymptotic &7.4&7.9&16.3\\
$\tilde{t}_{\mathcal{L}}(q=4)$&1.6&1.6&2.0&$\tilde{t}_{\mathcal{L}}(q=4)$&1.6&1.6&2.0\\
$\tilde{t}_{\mathcal{L}}(q=8)$&5.4&5.3&6.0&$\tilde{t}_{\mathcal{L}}(q=8)$&5.4&5.3&6.0\\
$\tilde{t}_{\mathcal{L}}(q=12)$&6.8&7.3&8.1&$\tilde{t}_{\mathcal{L}}(q=12)$&6.8&7.3&8.1\\
$\tilde{t}_{\mathcal{L}}(q=16)$&7.5&7.8&8.7&$\tilde{t}_{\mathcal{L}}(q=16)$&7.5&7.8&8.7\\
$\tilde{\tilde{t}}_{\mathcal{L}}(q=4)$&4.2&4.3&4.7&$\tilde{\tilde{t}}_{\mathcal{L}}(q=4)$&4.2&4.3&4.7\\
$\tilde{\tilde{t}}_{\mathcal{L}}(q=8)$&7.1&7.4&8.1&$\tilde{\tilde{t}}_{\mathcal{L}}(q=8)$&7.1&7.4&8.1\\
$\tilde{\tilde{t}}_{\mathcal{L}}(q=12)$&8.2&8.6&9.7&$\tilde{\tilde{t}}_{\mathcal{L}}(q=12)$&8.2&8.6&9.7\\
$\tilde{\tilde{t}}_{\mathcal{L}}(q=16)$&8.7&9.0&9.8&$\tilde{\tilde{t}}_{\mathcal{L}}(q=16)$&8.7&9.0&9.8\\
permutation&4.4&4.8&10.1&permutation&4.1&4.3&9.3\\
bootstrap&4.4&4.9&11.2&bootstrap&4.3&4.9&11.0\\
  \bottomrule
\end{tabularx}
\end{table}

\subsubsection{Inference in two-sample problem: Finite-sample power properties}

Next, we investigate finite-sample power properties of the tests considered. We report finite-sample size adjusted power for two-sample $t-$statistic and permutation tests.\footnote{Size adjustment is not performed for bootstrap tests as they are strongly dominated in terms of power by permutation test in all settings considered, see also \citet{Dufour}.}. Under size adjustment, the resulting empirical size of a given $t$-statistic-based robust test and its permutation counterpart coincide under the null hypothesis, thereby enabling meaningful power comparisons. 

We consider the following simulation designs.

Table \ref{tab_2_power_1} presents the size-adjusted power when the two samples come from different Singh-Maddala distributions $SM(a_0, b_0, c).$ The sample sizes are $N_1=N_2=200,$ and the number of groups is the same for $t-$statistic tests: $q_1=q_2=q.$ The first sample has a fixed Singh-Maddala distribution $SM(a_0, b_0, c_0)$ with $a_0=2.8,c_0=1.7$ and the corresponding tail index $\zeta_I=4.76$ and the distribution of the second sample varies, with $a_0=2.8$, $c=0.7,1.1,1.7,2.7,31.7$ and the corresponding tail indices $\zeta_Y=1.96, 3.08, 4.76, 7.56, 88.76.$ The permutation test appears to be the most powerful although the two-sample $t$-statistic tests (based on $\tilde{t}_{\mathcal{L}}$ in (\ref{Gen2})) have only slightly lower power (for $q=12,16$). The two-sample tests based on the $t-$statistic   $\tilde{t}_{\mathcal{L}}$ in (\ref{Gen2}) with $q_1=q_2=q$ are always more powerful than those based on the one-sample $t-$statistic $\tilde{\tilde{t}}_{\mathcal{L}}$ in (\ref{GenDiff}) in the differences of the group estimators with the same number of groups. In inference on both Theil and Gini indices, the power of $t-$statistic approaches based on (\ref{Gen2}) is very similar across $q=8, 12, 16$ if the second distribution is more light-tailed than the first one, so that $c>c_0$ and $\zeta_Y>\zeta_I$ and also very similar for $q=12, 16$ if the second distribution is more heavy-tailed than the first one, with $c<c_0$ and $\zeta_Y<\zeta_I.$ In the former case of more lighted second distribution ($c>c_0$ and $\zeta_Y>\zeta_I$), the best power is exhibited by $t-$statistic tests based on  (\ref{Gen2}) with $q_1=q_2=q=8, 12.$  In the latter case of more heavy-tailed second distribution ($c<c_0$ and $\zeta_Y<\zeta_I$), the most powerful $t-$statistic test for inference on Theil indices is the one based on  (\ref{Gen2}) with $q_1=q_2=16,$ and the second best test is the $t$-statistic test based on (\ref{Gen2}) with $q_1=q_2=12.$ Also, in the above case where the second distribution is more heavy-tailed than the first one, with $c<c_0$ and $\zeta_Y<\zeta_I,$ the most powerful $t-$statistic test for inference on Gini indices is the test based on  (\ref{Gen2}) with $q_1=q_2=12.$

\begin{table}[!h]
\centering
\small
\caption{Size-adjusted power, $\zeta_I=4.76,$ identical sample sizes $N_1=N_2=200$ \label{tab_2_power_1}}
\begin{tabularx}{0.95\textwidth}{cccccc|cccccc} \toprule
Theil\textbackslash $\zeta_Y$ &1.96&3.08&4.76&7.56&88.76& Gini \textbackslash $\zeta_Y $ &1.96&3.08&4.76&7.56&88.76\\\hline
asymptotic&87.2&35.6&4.7&23.6&90.4&asymptotic&97.5&39.2&4.5&23.3&91.4\\
$\tilde{t}_{\mathcal{L}}(q=4)$&62.6&22.5&4.7&17.6&72.3&$\tilde{t}_{\mathcal{L}}(q=4)$&82.4&26.2&4.5&17.5&75.4\\
$\tilde{t}_{\mathcal{L}}(q=8)$&82.7&29.4&4.7&20.7&83.6&$\tilde{t}_{\mathcal{L}}(q=8)$&93.6&31.6&4.5&19.6&84.1\\
$\tilde{t}_{\mathcal{L}}(q=12)$&88.3&30.5&4.7&21.4&83.5&$\tilde{t}_{\mathcal{L}}(q=12)$&94.7&32.5&4.5&20.3&83.6\\
$\tilde{t}_{\mathcal{L}}(q=16)$&91.1&30.8&4.7&20.8&82.8&$\tilde{t}_{\mathcal{L}}(q=16)$&94.7&31.3&4.5&19.2&81.9\\
$\tilde{\tilde{t}}_{\mathcal{L}}(q=4)$&43.6&17.3&4.7&15.2&55.2&$\tilde{\tilde{t}}_{\mathcal{L}}(q=4)$&63.4&20.0&4.5&14.7&57.9\\
$\tilde{\tilde{t}}_{\mathcal{L}}(q=8)$&75.5&26.3&4.7&19.5&77.2&$\tilde{\tilde{t}}_{\mathcal{L}}(q=8)$&89.2&28.8&4.5&18.5&77.7\\
$\tilde{\tilde{t}}_{\mathcal{L}}(q=12)$&83.6&27.5&4.7&19.3&78.7&$\tilde{\tilde{t}}_{\mathcal{L}}(q=12)$&92.0&29.2&4.5&18.1&78.5\\
$\tilde{\tilde{t}}_{\mathcal{L}}(q=16)$&88.4&28.7&4.7&19.6&79.4&$\tilde{\tilde{t}}_{\mathcal{L}}(q=16)$&93.1&29.8&4.5&18.2&78.8\\
permutation&91.6&37.4&4.7&21.9&88.7&permutation&97.6&39.7&4.5&21.6&90.0\\
bootstrap&77.7&33.6&4.3&20.6&87.2&bootstrap&95.2&39.0&4.6&21.8&90.2\\
  \bottomrule
\end{tabularx}
\end{table}

Table \ref{tab_2_power_2} provides the results on finite sample power properties of different inference approaches in the case of more heavy-tailed distributions. In the numerical analysis in the table, the fist sample has Singh-Maddala distribution $SM(a, b_0, c)$, with $a=2, c=1.1$ and $\zeta_I=2.2,$ and the second sample is from the Singh-Maddala distribution $SM(a, b_0, c)$ with $a=2,$ $c=0.7,0.9,1.1,1.5,3.7$ and the corresponding tail indices $\zeta_Y=1.4, 1.8, 2.2, 3, 7.4.$ The sample sizes are $N_1=N_2=200.$ According to the results in Table \ref{tab_2_power_2}, in the case of inference on Theil or Gini indices, the $t-$statistic tests based on $\tilde{t}_{\mathcal{L}}$ in (\ref{Gen2}) with $q_1=q_2=8, 12, 16$ are typically the most powerful (this is the case for not very lighted second distribution); in particular, they are typically more powerful than permutation tests. In the case of inference on Theil indices, the best power properties are exhibited by the $t-$statistic tests based on with $q=16,$ and the second best test is the $t$-statistic test based on (\ref{Gen2}) with $q_1=q_2=12.$ The choice of $q=12, 16$ also provides the best power properties for  $t-$statistic tests based on $\tilde{t}_{\mathcal{L}}$ in (\ref{Gen2}) in inference on Gini indices. 

\begin{table}[!h]
\centering
\small
\caption{Size-adjusted power, $\zeta_I=2.2,$ identical sample sizes $N_1=N_2=200$ \label{tab_2_power_2}}
\begin{tabularx}{0.95\textwidth}{cccccc|cccccc} \toprule
Theil\textbackslash $\zeta_Y $ &1.4&1.8&2.2&3&7.4&Gini\textbackslash $\zeta_Y $ &1.4&1.8&2.2&3&7.4\\\hline
asymptotic&49.76&13.71&4.71&17.72&85.99&asymptotic&59.93&15.61&4.75&21.05&95.03\\
$\tilde{t}_{\mathcal{L}}(q=4)$&28.37&9.25&4.71&15.83&73.85&$\tilde{t}_{\mathcal{L}}(q=4)$&39.79&10.82&4.75&17.89&83.77\\
$\tilde{t}_{\mathcal{L}}(q=8)$&40.72&11.3&4.71&18.76&87.21&$\tilde{t}_{\mathcal{L}}(q=8)$&51.02&12.83&4.75&20.52&92.41\\
$\tilde{t}_{\mathcal{L}}(q=12)$&45.08&11.97&4.71&19.95&90.05&$\tilde{t}_{\mathcal{L}}(q=12)$&52.19&13.36&4.75&21.2&92.33\\
$\tilde{t}_{\mathcal{L}}(q=16)$&47.94&12.4&4.71&20.1&91.22&$\tilde{t}_{\mathcal{L}}(q=16)$&52.51&13.08&4.75&20.71&91.51\\
$\tilde{\tilde{t}}_{\mathcal{L}}(q=4)$&20.04&7.72&4.71&12.81&56.42&$\tilde{\tilde{t}}_{\mathcal{L}}(q=4)$&28.35&9.3&4.75&14.38&64.9\\
$\tilde{\tilde{t}}_{\mathcal{L}}(q=8)$&34.93&9.77&4.71&16.56&80.37&$\tilde{\tilde{t}}_{\mathcal{L}}(q=8)$&44.19&11.25&4.75&18.05&86.69\\
$\tilde{\tilde{t}}_{\mathcal{L}}(q=12)$&40.99&10.53&4.71&18.18&86.32&$\tilde{\tilde{t}}_{\mathcal{L}}(q=12)$&48.82&12.43&4.75&19.97&89.29\\
$\tilde{\tilde{t}}_{\mathcal{L}}(q=16)$&45.19&11.98&4.71&19.17&89.14&$\tilde{\tilde{t}}_{\mathcal{L}}(q=16)$&50.57&12.85&4.75&20.14&89.97\\
permutation&34.17&11.79&4.71&17.33&89.95&permutation&48.24&13.6&4.75&19.63&94.18\\
bootstrap&26.18&8.92&3.32&13.78&78.16&bootstrap&43.1&12.25&3.87&18.39&91.1\\
  \bottomrule
\end{tabularx}
\end{table}

Tables \ref{tab_2_power_3} and \ref{tab_2_power_4} provide the results on finite-sample power properties of different inference approaches in the case of heavy-tailed distributions, including those considered in Table \ref{tab_2_power_2} ($a=2,$ $c=0.7,0.9,1.1,1.5$ and the corresponding tail indices $\zeta_Y=1.4, 1.8, 2.2, 3,$ and also $c=3.7,$ $\zeta_Y=7.4$ in the case of Theil indices and $c=2.2$ and $\zeta_Y=4.4$ in the case of Gini indices), and different sample sizes, with $N_1=200,$ $N_2=400$ in the former table and $N_1=400,$ $N_2=200$ in the later one. 

One can see that two-sample $t$-statistic tests are typically much more powerful than permutation tests if the more heavy-tailed distribution has larger sample size. Again, two-sample $t$-statistic tests based on $\tilde{t}_{\mathcal{L}}$ in (\ref{Gen2}) with the number of groups $q_1=q_2=q$ are always more powerful than those based on the one-sample $t-$statistic $\tilde{\tilde{t}}_{\mathcal{L}}$ in (\ref{GenDiff}) in the differences of the group estimators with the same number of groups.
 
\begin{table}[!h]
\centering
\footnotesize
\caption{Size-adjusted power, $\zeta_I=2.2,$ different sample sizes, $N_1=200$, $N_2=400$ \label{tab_2_power_3}}
\begin{tabularx}{0.95\textwidth}{cccccc|cccccc} \toprule
Theil\textbackslash $\zeta_Y$ &1.4&1.8&2.2& 3&7.4&Gini\textbackslash $\zeta_Y$ &1.4&1.8&2.2&3&4.4\\\hline
asymptotic&62.61&20.09&4.47&13.74&87.24&asymptotic&73.49&21.82&4.2&21.41&75.46\\
$\tilde{t}_{\mathcal{L}}(q=4)$&40.75&14.61&4.47&9.84&68.67&$\tilde{t}_{\mathcal{L}}(q=4)$&57.1&16.06&4.2&16.47&56.77\\
$\tilde{t}_{\mathcal{L}}(q=8)$&58.48&19.39&4.47&8.33&80.52&$\tilde{t}_{\mathcal{L}}(q=8)$&72.14&21.62&4.2&18.05&67.1\\
$\tilde{t}_{\mathcal{L}}(q=12)$&65.39&22.06&4.47&5.08&78.17&$\tilde{t}_{\mathcal{L}}(q=12)$&75.29&22.9&4.2&15.64&63.4\\
$\tilde{t}_{\mathcal{L}}(q=16)$&70.03&23.41&4.47&3.06&73.4&$\tilde{t}_{\mathcal{L}}(q=16)$&78.8&24.63&4.2&14.53&62.81\\
$\tilde{\tilde{t}}_{\mathcal{L}}(q=4)$&28.92&11.94&4.47&7.94&49.73&$\tilde{\tilde{t}}_{\mathcal{L}}(q=4)$&39.41&12.4&4.2&11.21&36.8\\
$\tilde{\tilde{t}}_{\mathcal{L}}(q=8)$&51.49&17.78&4.47&7.41&72.84&$\tilde{\tilde{t}}_{\mathcal{L}}(q=8)$&64.51&19.31&4.2&14.92&57.43\\
$\tilde{\tilde{t}}_{\mathcal{L}}(q=12)$&59.87&20.39&4.47&4.48&71.01&$\tilde{\tilde{t}}_{\mathcal{L}}(q=12)$&72.52&22.26&4.2&15.32&60.2\\
$\tilde{\tilde{t}}_{\mathcal{L}}(q=16)$&65.75&22.16&4.47&2.96&68.15&$\tilde{\tilde{t}}_{\mathcal{L}}(q=16)$&75.23&22.96&4.2&13.53&58.66\\
permutation&45.47&13.75&4.47&23.59&96.1&permutation&58.98&16.22&4.2&27.71&82.51\\
bootstrap&28.62&9.72&3.23&20.1&89.22&bootstrap&50.06&14.59&3.85&25.86&78.94\\
  \bottomrule
\end{tabularx}
\end{table}

\begin{table}[!h]
\centering
\footnotesize
\caption{Size-adjusted power, $\zeta_I=2.2,$ different sample sizes, $N_1=400$, $N_2=200$  \label{tab_2_power_4}}
\begin{tabularx}{0.91\textwidth}{cccccc|cccccc} \toprule
Theil\ $\zeta_Y$ &1.4 &1.8&2.2&3&7.4&Gini\ $\zeta_Y$ &1.4&1.8&2.2&3&4.4\\\hline
asymptotic&44.56&9.2&4.15&27.76&92.67&asymptotic&64.67&14.91&4.45&32.15&84.15\\
$\tilde{t}_{\mathcal{L}}(q=4)$&19.25&4.87&4.15&23.21&83.34&$\tilde{t}_{\mathcal{L}}(q=4)$&38.92&9.19&4.45&25.61&70.46\\
$\tilde{t}_{\mathcal{L}}(q=8)$&21.13&3.53&4.15&28.2&92.52&$\tilde{t}_{\mathcal{L}}(q=8)$&46.64&9.29&4.45&31.18&80.64\\
$\tilde{t}_{\mathcal{L}}(q=12)$&15.17&2.01&4.15&27.94&93.97&$\tilde{t}_{\mathcal{L}}(q=12)$&44&7.99&4.45&31.57&80.67\\
$\tilde{t}_{\mathcal{L}}(q=16)$&11.7&1.3&4.15&30.44&95.66&$\tilde{t}_{\mathcal{L}}(q=16)$&42.78&7.21&4.45&32.73&81.5\\
$\tilde{\tilde{t}}_{\mathcal{L}}(q=4)$&12.9&4.23&4.15&17.61&65.25&$\tilde{\tilde{t}}_{\mathcal{L}}(q=4)$&26.82&7.66&4.45&19.65&52.38\\
$\tilde{\tilde{t}}_{\mathcal{L}}(q=8)$&16.79&3.02&4.15&24.4&86.85&$\tilde{\tilde{t}}_{\mathcal{L}}(q=8)$&41.27&8.31&4.45&28.27&74.04\\
$\tilde{\tilde{t}}_{\mathcal{L}}(q=12)$&13.08&1.77&4.15&25.76&90.44&$\tilde{\tilde{t}}_{\mathcal{L}}(q=12)$&40.46&7.34&4.45&29.29&76.66\\
$\tilde{\tilde{t}}_{\mathcal{L}}(q=16)$&10.53&1.23&4.15&28.23&93.5&$\tilde{\tilde{t}}_{\mathcal{L}}(q=16)$&40.28&7.06&4.45&30.68&78.65\\
permutation&44.57&14.95&4.15&21.71&97.47&permutation&62.78&18.55&4.45&24.75&77.41\\
bootstrap&40.34&12.9&3.58&16.33&84.87&bootstrap&59.58&17.66&4.12&23.13&74.1\\
  \bottomrule
\end{tabularx}
\end{table}

Table \ref{tab_2_power_5} gives the results on finite-sample size adjusted power of different inference approaches in the same same distributional settings as in Table \ref{tab_2_power_3} and sample sizes $N_1=200$ and $N_2=800.$ Similarly, \ref{tab_2_power_6} provides the results on finite-sample size adjusted power properties of the approaches in the same settings as in Table \ref{tab_2_power_4} and sample sizes $N_1=800$ and $N_2=200$. We also consider different combinations of (not necessarily equal) numbers $q_1$ and $q_2$ of groups for $t-$statistic inference approaches. According to the results in Tables \ref{tab_2_power_5} and \ref{tab_2_power_6}, if smaller sample is more heavy-tailed then the power of all two-sample $t$-statistic tests is dominated by that of permutation tests. Otherwise, if the larger sample is more heavy-tailed then the power properties of two-sample $t$-statistics tests (except the tests with very small $q_1$ and $q_2$) are typically considerably better than those of permutation test. One can further see that for inference on Theil indices, the best (compromise) choice of the number of groups in $t-$statistic testing approaches will be $q_1=12$, $q_2=6$ or \textit{vice versa} because this choice leads to correct size and good power in comparison to other size-controlled two-sample $t$-statistic tests. For Gini indices, the finite-sample power properties are not very sensitive to choice $q_1$ and $q_2$. Interestingly, even if the samples differ 4 times as in the tables, the choice $q_1=q_2=8, 12, 16$ leads to a very good size adjusted power and seems to be one of the best across all combinations of $q_1$ and $q_2$. The choice $q_1=12$ and $q_2=9$ also a good choice and leads to power properties of $t-$statistic inference approaches that are comparable or slightly better than in the case $q_1=q_2=8, 12, 16$. The choice of the different number of groups $q_1$ and $q_2$ may be useful if the sizes of two samples differ very much.

\begin{table}[!ht]
\centering
\footnotesize
\caption{Size-adjusted power, $\zeta_I=2.2,$ different sample sizes, $N_1=200$, $N_2=800$ \label{tab_2_power_5}}
\begin{tabularx}{0.95\textwidth}{cccccc|cccccc} \toprule
Theil \textbackslash $\zeta_Y$ &1.4&1.8&2.2&3&7.4&Gini \textbackslash $\zeta_Y$ &1.4&1.8&2.2&3&4.4\\\hline
asymptotic&64.9&23.7&4.7&7.6&82.03&asymptotic&81.2&29.1&4.8&21.8&79.6\\
$\tilde{t}_{\mathcal{L}}(q=4)$&45.8&18.5&4.7&6.2&58.53&$\tilde{t}_{\mathcal{L}}(q=4)$&69.9&22.6&4.8&16.8&58.7\\
$\tilde{t}_{\mathcal{L}}(q=8)$&58.8&21.4&4.7&2.3&60.79&$\tilde{t}_{\mathcal{L}}(q=8)$&83.0&28.0&4.8&15.0&65.3\\
$\tilde{t}_{\mathcal{L}}(q=12)$&65.5&22.5&4.7&0.8&49.03&$\tilde{t}_{\mathcal{L}}(q=12)$&87.1&30.6&4.8&11.6&60.9\\
$\tilde{t}_{\mathcal{L}}(q=16)$&71.4&24.4&4.7&0.3&37.4&$\tilde{t}_{\mathcal{L}}(q=16)$&89.7&31.8&4.8&9.0&57.1\\
$\tilde{t}_{\mathcal{L}}(q_1=4,q_2=3)$&41.9&16.8&4.7&8.6&56.32&$\tilde{t}_{\mathcal{L}}(q_1=4,q_2=3)$&62.8&19.6&4.8&17.5&54.6\\
$\tilde{t}_{\mathcal{L}}(q_1=8,q_2=6)$&58.9&21.2&4.7&4.1&64.72&$\tilde{t}_{\mathcal{L}}(q_1=8,q_2=6)$&80.0&27.2&4.8&17.6&66.4\\
$\tilde{t}_{\mathcal{L}}(q_1=12,q_2=9)$&67.1&22.8&4.7&2.2&62.79&$\tilde{t}_{\mathcal{L}}(q_1=12,q_2=9)$&84.6&28.9&4.8&15.6&66.8\\
$\tilde{t}_{\mathcal{L}}(q_1=16,q_2=12)$&72.3&25.0&4.7&0.9&55.22&$\tilde{t}_{\mathcal{L}}(q_1=16,q_2=12)$&87.2&30.6&4.8&13.5&64.5\\
$\tilde{t}_{\mathcal{L}}(q_1=4,q_2=2)$&29.1&11.4&4.7&13.9&52.99&$\tilde{t}_{\mathcal{L}}(q_1=4,q_2=2)$&45.7&13.6&4.8&19.0&49.1\\
$\tilde{t}_{\mathcal{L}}(q_1=8,q_2=4)$&57.5&20.1&4.7&8.6&66.29&$\tilde{t}_{\mathcal{L}}(q_1=8,q_2=4)$&74.3&23.6&4.8&19.5&63.5\\
$\tilde{t}_{\mathcal{L}}(q_1=12,q_2=6)$&66.1&22.4&4.7&5.7&69.66&$\tilde{t}_{\mathcal{L}}(q_1=12,q_2=6)$&79.3&26.1&4.8&19.2&68.1\\
$\tilde{t}_{\mathcal{L}}(q_1=16,q_2=8)$&71.4&23.9&4.7&4.2&70.21&$\tilde{t}_{\mathcal{L}}(q_1=16,q_2=8)$&81.8&27.0&4.8&18.3&69.7\\
permutation&54.4&16.3&4.7&33.6&97.68&permutation&65.8&18.8&4.8&38.9&91.6\\
bootstrap&30.3&10.3&3.6&29.7&95.78&bootstrap&56.0&16.5&4.4&36.9&90.2\\
  \bottomrule
\end{tabularx}
\end{table}

\begin{table}[!ht]
\centering
\footnotesize
\caption{Size-adjusted power, $\zeta_I=2.2,$ different sample sizes, $N_1=800$, $N_2=200$  \label{tab_2_power_6}}
\begin{tabularx}{0.95\textwidth}{cccccc|cccccc} \toprule
Theil \textbackslash $\zeta_Y$ &1.4&1.8&2.2&3&7.4&Gini \textbackslash $\zeta_Y$ &1.4&1.8&2.2&3&4.4\\\hline
asymptotic&34.0&4.8&4.9&35.8&93.74&asymptotic&65.1&11.7&4.8&40.8&91.4\\
$\tilde{t}_{\mathcal{L}}(q=4)$ &11.2&2.6&4.9&28.3&87.08&$\tilde{t}_{\mathcal{L}}(q=4)$ &35.4&7.1&4.8&32.2&81.6\\
$\tilde{t}_{\mathcal{L}}(q=8)$ &7.3&1.1&4.9&33.8&92.78&$\tilde{t}_{\mathcal{L}}(q=8)$ &41.1&6.2&4.8&40.3&90.1\\
$\tilde{t}_{\mathcal{L}}(q=12)$ &3.2&0.6&4.9&34.5&94.11&$\tilde{t}_{\mathcal{L}}(q=12)$ &37.0&4.4&4.8&42.4&91.0\\
$\tilde{t}_{\mathcal{L}}(q=16)$ &1.2&0.5&4.9&35.3&94.63&$\tilde{t}_{\mathcal{L}}(q=16)$ &30.6&2.8&4.8&41.1&90.1\\
$\tilde{t}_{\mathcal{L}}(q_1=4,q_2=3)$&14.9&4.1&4.9&26.7&85.5&$\tilde{t}_{\mathcal{L}}(q_1=4,q_2=3)$&35.2&8.6&4.8&29.8&76.5\\
$\tilde{t}_{\mathcal{L}}(q_1=8,q_2=6)$&11.3&1.8&4.9&33.4&93.34&$\tilde{t}_{\mathcal{L}}(q_1=8,q_2=6)$&43.2&7.5&4.8&38.3&88.6\\
$\tilde{t}_{\mathcal{L}}(q_1=12,q_2=9)$&8.4&1.0&4.9&35.7&95.69&$\tilde{t}_{\mathcal{L}}(q_1=12,q_2=9)$&45.6&6.4&4.8&41.9&91.2\\
$\tilde{t}_{\mathcal{L}}(q_1=16,q_2=12)$&4.3&0.6&4.9&36.2&96.56&$\tilde{t}_{\mathcal{L}}(q_1=16,q_2=12)$&40.3&5.1&4.8&41.4&90.9\\
$\tilde{t}_{\mathcal{L}}(q_1=4,q_2=2)$&20.6&7.4&4.9&20.0&77.76&$\tilde{t}_{\mathcal{L}}(q_1=4,q_2=2)$&32.6&9.9&4.8&21.2&60.0\\
$\tilde{t}_{\mathcal{L}}(q_1=8,q_2=4)$&17.3&3.9&4.9&33.5&94.34&$\tilde{t}_{\mathcal{L}}(q_1=8,q_2=4)$&42.7&9.3&4.8&35.1&85.7\\
$\tilde{t}_{\mathcal{L}}(q_1=12,q_2=6)$&15.3&2.4&4.9&35.9&96.79&$\tilde{t}_{\mathcal{L}}(q_1=12,q_2=6)$&47.6&8.7&4.8&38.5&89.2\\
$\tilde{t}_{\mathcal{L}}(q_1=16,q_2=8)$&13.5&1.8&4.9&37.4&97.96&$\tilde{t}_{\mathcal{L}}(q_1=16,q_2=8)$&49.1&8.1&4.8&40.1&90.6\\
permutation&53.5&17.2&4.9&27.4&99.34&permutation&73.7&22.5&4.8&30.5&86.8\\
bootstrap&53.4&16.2&4.0&20.7&90.28&bootstrap&73.6&21.5&4.4&28.9&84.5\\
  \bottomrule
\end{tabularx}
\end{table}

Summarizing the results, the two-sample $t$-statistic robust approaches to testing equality of two inequality measures or inference on their difference appear to be useful complements to other inference methods, including computationally expensive bootstrap and permutation-based inference methods. Finite-sample properties of the $t-$statistic inference approaches appear to be better in the case of testing equality and comparisons of Gini measures as compared to the case of the Theil measures as the former measures are more robust to heavy tails.

In applications of two-sample $t$-statistic inference approaches, the appropriate choice of the numbers $q_1$ and $q_2$ of groups is needed. The most simple way to choose the numbers of groups in the case of distributions that are not very different from each other is to have $q_1/q_2$ (approximately) equal to $N_1/N_2$ so that the sizes of all the groups considered are the same. If two distributions have similar tail indices, then in the case of inference on Gini measures, $q_1$ and $q_2$ may be taken to be equal. In general, the size of the groups in the sample from a more heavy-tailed distribution should be larger than the size of the groups from a less heavy-tailed distribution. E.g., in the case of equally sized samples, one should take the number of groups in the more heavy-tailed sample to be less than the number of groups in the less heavy-tailed sample. 
 
\section{ Empirical application: Income inequality across Russian regions}\label{emp}

This section presents empirical results on comparisons of Gini coefficients in Moscow and Russian regions using the asymptotic, permutation, bootstrap and the robust $t-$statistic inference approaches considered in this paper. 

The empirical analysis is based on a large database on the results of household income surveys conducted by the Federal State Statistics Service of Russia (Rosstat) in 2017 (available at $https://www.gks.ru/free\_doc/new\_site/vndn-2017/index.html$;\\ $https://www.gks.ru/free\_doc/new\_site/vndn-2017/OHousehold.html$). The database covers 160,000 households in Russian regions, and provides the data on, among many other variables, households' total income. The analysis of  income inequality indices and their comparisons in this section is based on the above income levels of Russian household normalized, following Rosstat's methodology, by the total number of households' members.

Table A.1 in the appendix provides the $p-$values for the above tests of the null hypothesis $H_0: G_M=G_R$ against the alternative $H_a: G_M\neq G_R,$ where $G_M$ is the Gini coefficient in Moscow and $G_R$ is the Gini coefficient in Russian region $R.$ The entries in the table in bold are the $p-$values not greater than 0.05.

The table also provides the values of the Gini coefficients and the (bias-corrected) log-log rank-size regression estimates (with 5\% tail truncation) of tail indices $\zeta$ of the income distribution among $N_2$ households surveyed in the regions (see \citeauthor{GI}, \citeyear{GI}). It also provides the values of the ratio $N_1/N_2,$ where $N_1$ is the number of households surveyed in Moscow. It should be noted that if $q_1$ or $q_2> 14$, we can use only the significance level less than 0.083.

The Gini coefficients in Moscow and Russian regions range from 0.236  (Tambov Region) to 0.354 (the Republic of Ingushetia) indicating low to moderate inequality; Tyva Republic has the Gini coefficient of 0.423 (Tyva Republic) indicating high inequality. The value of the Gini coefficient for Moscow is 0.264 indicating rather low inequality.

The point log-log rank-size regression estimates $\zeta$ of tail indices of income distribution in most of Russian regions lie in the interval $(3, 6),$ with the exception of Karachay-Cherkess ($\zeta=2.08$) and Mari El ($\zeta=2.29$) Republics and Krasnodar ($\zeta=2.6$), Kursk ($\zeta=2.75$) and Tyumen ($\zeta=2.71$) regions. The corresponding confidence intervals for tail indices of income distribution in most of Russian regions lie on the right of 2 implying finite second moments and finite variances.  The 95\% confidence intervals for tail indices of income distribution in Krasnodar, Krasnoyarsk, Stavropol, Khabarovsk, Arkhangelsk, Astrakhan, Belgorod, Vladimir, Volgograd, Vologda, Voronezh, Ivanovo, Tver, Kemerovo, Kurgan, Kursk, Lipetsk, Magadan, Murmansk, Novosibirsk, Omsk, Oryol, Penza, Pskov, Ryazan, Sakhalin, Sverdlovsk, Smolensk, Tambov, Tomsk, Tyumen, Ulyanovsk and Yaroslav regions; Altai, Buryatia, Ingushetia, Kabardino-Balkar, Kalmykia, Karachay-Cherkess, Karelia, Komi, Mari El, Mordovia, North Osetia, Tyva and Sakha Republics; Chukotka, Khanty-Mansi and Nenets Autonomous Districts and Kamchatka Kray intersect with the interval $(1.5, 3)$ where tail indices of income distribution in developed countries typically lie.  
The 95\% confidence intervals for tail indices of income distribution in Amur, Bryansk, Chelyabinsk, Irkutsk, Kaliningrad, Kaluga, Kirov, Kostroma, Leningrad, Moscow (the tail index estimate is 3,96 with the 95\% confidence interval $(3.44, 4.48)$), Nizhny Novgorod, Novgorod, Orenburg, Perm, Rostov, Samara, Saratov, Sevastopol and Tula regions; Adygeya, Bashkortostan, Chuvash, Crimea, Dagestan, Khakassia and Tatarstan Republics and Kamchatka, Primorsky and Zabaykalsky Krays lie on the right of 3 thus implying finite third moments and variances.

According to the table, on the base of all the tests considered, including the $t-$statistic tests with most of the values $q_1, q_2,$ the null hypothesis $H_0: G_M=G_R$ is rejected in favor of the alternative $H_a: G_M>G_R$ (at the level 2.5\%) for the Republic of Tatarstan,  Sevastopol City and Bryansk, Kostroma,  Tambov and Tula regions. For Penza, Smolensk and Ulyanovsk regions and Udmurtia, $H_0: G_M=G_R$ is rejected in favor of $H_a: G_M>G_R$ on the base of the asymptotic, bootstrap, permutation and the $t-$statistic tests with some of the values $q_1, q_2$ in the table.

Further, according to all the tests considered, including the $t-$statistic tests for most of the values $q_1, q_2,$ the null hypothesis $H_0: G_M=G_R$ is rejected in favor of the alternative $H_a: G_M<G_R$ (at the level 2.5\%)  for Amur, Chelyabinsk, Irkutsk, Khabarovsk, Krasnodar, Krasnoyarsk, Kurgan, Moscow, Sakhalin and Jewish and Yamalo-Nenets Autonomous regions as well as for the Republics of Bashkortostan,  Buryatia, Dagestan, Ingushetia, Kalmykia, Khakassia and Sakha (Yakutia); Altai, Chechen, Kabardino-Balkar, Karachay-Cherkess, Komi and Tyva Republics; Kamchatka, Primorskiy, Zabaykalsky Krays; Khanty-Mansi and Nenets Autonomous Okrugs and Chukotka Autonomous District. For Astrakhan, Kaliningrad, Kemerovo, Novosibirsk, Omsk, Penza, Smolensk, Sverdlovsk, Tomsk and Tyumen Regions, $H_0: G_M=G_R$ is rejected in favor of $H_a: G_M<G_R$ on the base of the asymptotic, bootstrap, permutation and the $t-$statistic tests for some of the values $q_1, q_2$ in the table.

Two conclusions are interesting to note. 

First, income inequality appears to be higher in most of the Russian Regions as compared to Moscow. 

Second, the conclusions of all the approaches to testing equality of the Gini coefficients $G_M$ and $G_R$ considered - the asymptotic, bootstrap, permutation and the robust $t-$statistic tests - for the above regions  agree among themselves. Two exceptions are Belgorod and Novgorod Regions, where $H_0: G_M=G_R$ is not rejected in favor of $H_a: G_M<G_R$ on the base of the asymptotic, bootstrap, permutation, but is rejected on the base of robust $t-$statistic tests for some values of $q_1, q_2.$ 

\section{Conclusion and suggestions for further research}\label{conclude}

Empirical analyses on inequality measurement and those in other fields in economics and finance often face the difficulty that the data is correlated, heterogeneous or heavy-tailed in some unknown fashion. In particular, as has been documented in numerous studies, observations on many variables of interest, including income, wealth and financial returns, typically exhibit heterogeneity, dependence and heavy tails in the form of commonly observed Pareto or power laws.

The paper focuses on applications of the recently developed \textit{t}-statistic based robust inference approaches in the analysis of inequality measures and their comparisons under the above problems. Following the approaches, in particular, a robust large sample test on equality of two parameters of interest (e.g., a test of equality of inequality measures in two regions or countries considered) is conducted as follows: The data in the two samples dealt with is partitioned into fixed numbers $q_1, q_2\ge 2$ (e.g., $q_1=q_2=2, 4, 8$) of groups, the parameters (inequality measures dealt with) are estimated for each group, and inference is based on a standard two-sample $t-$test with the resulting $q_1, q_2$ group estimators. Robust $t-$statistic approaches result in valid inference under general conditions that group estimators of parameters (e.g., inequality measures) considered are asymptotically independent, unbiased and Gaussian of possibly different variances, or weakly converge, at an arbitrary rate, to independent scale mixtures of normal random variables. These conditions are typically satisfied in empirical applications even under pronounced heavy-tailedness and heterogeneity and possible dependence in observations. 

The methods dealt with in the paper complement and compare favorably with other inference approaches available in the literature. We illustrate application of the proposed robust inference approaches by an empirical analysis of income inequality measures and their comparisons across different regions in Russia. 

The $t-$statistic robust inference approaches, including the two-sample approaches for inference on equality of and the difference between parameters of interest considered in this paper are simple to use and have a wide range of applicability in econometric and statistical analysis under the problems of heterogeneity, dependence and heavy-tailedness in observations. The approaches do not require at all estimation of limiting variances of estimators of interest, in contrast to inference methods based on consistent, e.g., HAC or clustered, standard errors that often have pure finite sample properties, especially under pronounced heterogeneity and dependence in observations. In addition, the inference approaches can be used under extremes and outliers in observations generated by heavy-tailedness with infinite variances and also in settings where observations (e.g., on income or wealth levels) in each of the samples considered are \emph{dependent} among themselves - for instance, due to spatial or clustered dependence, common shocks affecting them, or, in the case of time series or panel data on income or wealth levels, due to autocorrelation and dependence in observations over time. Further, in the case of testing for equality of inequality measures or inference on their difference in two populations using two samples of possibly dependent observations, as above, the $t-$statistic inference approaches may be used under \emph{an arbitrary} dependence \emph{between} the samples as well as under possibly \emph{unequal} sample sizes. 

In addition to inference on inequality and wealth indices dealt with in this work, the approaches may also be applied in inference on and comparisons of poverty and concentration indices where, as is well-known, the presence of extreme values, outliers, heavy-tailedness and heterogeneity makes problematic their applicability and the use of asymptotic methods in inference on the indices similar to the case of inequality measures (see, among others, Appendix B.1 in Section E7 in \citeauthor{Mand}, \citeyear{Mand}, \citeauthor{DF}, \citeyear{DF}, and Section 3.3.2 in \citeauthor{IIW}, \citeyear{IIW}) as well as in inference on tail indices in power laws (\ref{power1}) for income and wealth distributions and corresponding measures of top inequality (see the discussion in Section \ref{num} and references therein). These and other applications of the $t-$statistic robust inference approaches are currently under way by the authors and their co-authors.

\small{
\bibliographystyle{agsm}
\bibliography{bib}}
\noindent \textbf{}

\newpage
\appendix {\centerline{\textbf{\LARGE{Appendix A: Tables}}}} \label{AppA}
\begin{landscape}
\begin{table}[!h]
\centering
\scriptsize
\caption{Empirical results: $p-$values, Gini measure\label{tab_emp_1}}
\begin{tabularx}{1.12\textwidth}{ccccccccccccccccccccccccccccc} \toprule
&\begin{sideways}Altai region\end{sideways}&\begin{sideways}Krasnodar region\end{sideways}&\begin{sideways}Krasnoyarsk region\end{sideways}&\begin{sideways}Primorsky Krai\end{sideways}&\begin{sideways}Stavropol region\end{sideways}&\begin{sideways}Khabarovsk region\end{sideways}&\begin{sideways}Amur region\end{sideways}&\begin{sideways}Arkhangelsk region\end{sideways}&\begin{sideways}Astrakhan region\end{sideways}&\begin{sideways}Nenets Autonomous Okrug\end{sideways}&\begin{sideways}Belgorod region\end{sideways}&\begin{sideways}Bryansk region\end{sideways}&\begin{sideways}Vladimir region\end{sideways}&\begin{sideways}Volgograd region\end{sideways}&\begin{sideways}Vologda Region\end{sideways}&\begin{sideways}Voronezh region\end{sideways}&\begin{sideways}Nizhny Novgorod Region\end{sideways}\\\hline
Gini&0.273&0.291&0.327&0.284&0.269&0.315&0.300&0.274&0.280&0.322&0.253&0.243&0.252&0.256&0.261&0.258&0.265\\
$N_2$&2568&4392&2832&2160&2400&1608&1296&1488&1416&480&1752&1584&1752&2520&1512&2496&3360\\
$N_1/N_2$&3.50&2.05&3.18&4.17&3.75&5.60&6.94&6.05&6.36&18.75&5.14&5.68&5.14&3.57&5.95&3.61&2.68\\\hline
$\zeta$&3.50&2.60&3.03&4.31&3.93&3.99&5.04&4.10&4.37&4.77&3.74&4.60&3.67&3.57&3.92&3.76&3.93\\\hline
asymptotic&0.18&\textbf{0.00}&\textbf{0.00}&\textbf{0.00}&0.44&\textbf{0.00}&\textbf{0.00}&0.14&\textbf{0.02}&\textbf{0.00}&0.08&\textbf{0.00}&0.07&0.16&0.70&0.28&0.92\\
$q_1=q_2=4$&0.25&\textbf{0.04}&\textbf{0.00}&0.06&0.52&\textbf{0.00}&\textbf{0.02}&0.38&0.08&\textbf{0.04}&0.31&0.07&0.31&0.38&0.73&0.28&0.92\\
$q_1=q_2=8$&0.30&\textbf{0.02}&\textbf{0.00}&\textbf{0.01}&0.50&\textbf{0.00}&\textbf{0.00}&0.34&\textbf{0.02}&\textbf{0.00}&0.22&\textbf{0.02}&0.16&0.23&0.69&0.42&0.96\\
$q_1=q_2=12$&0.27&\textbf{0.01}&\textbf{0.00}&\textbf{0.01}&0.43&\textbf{0.00}&\textbf{0.00}&0.28&\textbf{0.02}&\textbf{0.00}&0.17&\textbf{0.00}&0.14&0.15&0.69&0.33&0.98\\
$q_1=q_2=16$&0.15&\textbf{0.00}&\textbf{0.00}&\textbf{0.01}&0.35&\textbf{0.00}&\textbf{0.00}&0.31&\textbf{0.02}&\textbf{0.00}&\textbf{0.02}&\textbf{0.00}&0.07&0.11&0.63&0.28&0.97\\
$q_1=4,q_2=3$&0.33&0.13&\textbf{0.02}&\textbf{0.03}&0.57&\textbf{0.01}&\textbf{0.03}&0.50&0.22&\textbf{0.04}&0.37&0.06&0.35&0.34&0.68&0.29&0.90\\
$q_1=8,q_2=6$&0.34&\textbf{0.03}&\textbf{0.00}&\textbf{0.03}&0.40&\textbf{0.00}&\textbf{0.00}&0.31&0.09&\textbf{0.01}&0.26&\textbf{0.01}&0.19&0.23&0.69&0.22&0.95\\
$q_1=12,q_2=9$&0.23&\textbf{0.01}&\textbf{0.00}&\textbf{0.05}&0.30&\textbf{0.00}&\textbf{0.00}&0.23&\textbf{0.03}&\textbf{0.00}&\textbf{0.03}&\textbf{0.02}&0.17&0.19&0.71&0.37&0.97\\
$q_1=16,q_2=12$&0.27&\textbf{0.01}&\textbf{0.00}&\textbf{0.01}&0.43&\textbf{0.00}&\textbf{0.00}&0.28&\textbf{0.02}&\textbf{0.00}&0.16&\textbf{0.00}&0.13&0.15&0.68&0.32&1.00\\
$q_1=4,q_2=2$&0.48&0.20&0.10&0.19&0.61&0.09&0.18&0.46&0.15&0.12&0.55&0.11&0.21&0.60&0.60&0.47&0.95\\
$q_1=8,q_2=4$&0.22&\textbf{0.04}&\textbf{0.00}&0.05&0.50&\textbf{0.00}&\textbf{0.02}&0.37&0.07&\textbf{0.04}&0.3&0.07&0.30&0.36&0.71&0.24&0.92\\
$q_1=12,q_2=6$&0.33&\textbf{0.03}&\textbf{0.00}&\textbf{0.02}&0.38&\textbf{0.00}&\textbf{0.00}&0.30&0.08&\textbf{0.01}&0.26&\textbf{0.01}&0.19&0.22&0.68&0.20&0.94\\
$q_1=16,q_2=8$&0.29&\textbf{0.02}&\textbf{0.00}&\textbf{0.01}&0.49&\textbf{0.00}&\textbf{0.00}&0.33&\textbf{0.02}&\textbf{0.00}&0.21&\textbf{0.02}&0.16&0.21&0.68&0.41&0.96\\
$q_1=8,q_2=2$&0.47&0.19&0.09&0.18&0.60&0.09&0.18&0.45&0.13&0.12&0.55&0.09&0.18&0.60&0.54&0.45&0.95\\
$q_1=12,q_2=3$&0.29&0.13&\textbf{0.02}&\textbf{0.01}&0.54&\textbf{0.01}&\textbf{0.02}&0.49&0.21&\textbf{0.04}&0.35&0.05&0.34&0.31&0.64&0.21&0.89\\
$q_1=16,q_2=4$&0.21&\textbf{0.04}&\textbf{0.00}&\textbf{0.05}&0.48&\textbf{0.00}&\textbf{0.02}&0.37&0.07&\textbf{0.04}&0.29&0.06&0.29&0.35&0.70&0.21&0.92\\
permutation&0.15&\textbf{0.01}&\textbf{0.00}&\textbf{0.00}&0.48&\textbf{0.00}&\textbf{0.00}&0.15&\textbf{0.02}&\textbf{0.00}&0.08&\textbf{0.00}&0.08&0.19&0.72&0.26&0.92\\
bootstrap&0.20&\textbf{0.00}&\textbf{0.00}&\textbf{0.00}&0.43&\textbf{0.00}&\textbf{0.00}&0.15&\textbf{0.02}&\textbf{0.00}&0.08&\textbf{0.00}&0.10&0.17&0.73&0.33&0.87\\
  \bottomrule
\end{tabularx}
\end{table}
\end{landscape}

\begin{landscape}
\begin{table}[!h]
\centering
\scriptsize
\captcont*{Empirical results: p-values, Gini measure, Continued\label{tab_emp_2}}
\begin{tabularx}{1.12\textwidth}{ccccccccccccccccccccccccccccc} \toprule
&\begin{sideways}Ivanovo region\end{sideways}&\begin{sideways}Irkutsk region\end{sideways}&\begin{sideways}The Republic of Ingushetia\end{sideways}&\begin{sideways}Kaliningrad region\end{sideways}&\begin{sideways}Tver region\end{sideways}&\begin{sideways}Kaluga region\end{sideways}&\begin{sideways}Kamchatka Krai\end{sideways}&\begin{sideways}Kemerovo region\end{sideways}&\begin{sideways}Kirov region\end{sideways}&\begin{sideways}Kostroma region\end{sideways}&\begin{sideways}Republic of Crimea\end{sideways}&\begin{sideways}Samara Region\end{sideways}&\begin{sideways}Kurgan region\end{sideways}&\begin{sideways}Kursk region\end{sideways}&\begin{sideways}Saint Petersburg city\end{sideways}&\begin{sideways}Leningrad region\end{sideways}&\begin{sideways}Lipetsk region\end{sideways}\\\hline
Gini&0.256&0.309&0.354&0.276&0.260&0.256&0.300&0.273&0.259&0.247&0.268&0.268&0.287&0.262&0.264&0.269&0.250\\
$N_2$&1512&2448&600&1368&1704&1440&888&2808&1680&1248&1992&3168&1368&1488&4344&1944&1488\\
$N_1/N_2$&5.95&3.68&15.00&6.58&5.28&6.25&10.14&3.21&5.36&7.21&4.52&2.84&6.58&6.05&2.07&4.63&6.05\\\hline
$\zeta$&3.77&4.51&3.69&5.05&3.94&6.00&3.34&3.76&5.15&4.77&4.45&4.49&3.18&2.75&4.27&5.63&3.67\\\hline
asymptotic&0.2&\textbf{0.00}&\textbf{0.00}&\textbf{0.05}&0.48&0.17&\textbf{0.00}&0.12&0.42&\textbf{0.01}&0.51&0.41&\textbf{0.01}&0.86&0.98&0.39&0.06\\
$q_1=q_2=4$&0.24&\textbf{0.01}&\textbf{0.01}&0.10&0.59&0.36&\textbf{0.02}&0.12&0.38&0.06&0.53&0.54&0.05&0.80&0.97&0.53&0.30\\
$q_1=q_2=8$&0.14&\textbf{0.00}&\textbf{0.00}&\textbf{0.02}&0.57&0.23&\textbf{0.01}&0.10&0.36&\textbf{0.02}&0.38&0.57&\textbf{0.02}&0.76&0.96&0.37&0.12\\
$q_1=q_2=12$&0.11&\textbf{0.00}&\textbf{0.00}&0.07&0.50&0.23&\textbf{0.03}&\textbf{0.03}&0.38&\textbf{0.01}&0.57&0.47&\textbf{0.02}&0.69&0.94&0.24&0.09\\
$q_1=q_2=16$&0.1&\textbf{0.00}&\textbf{0.00}&\textbf{0.02}&0.46&0.15&\textbf{0.00}&0.10&0.35&\textbf{0.01}&0.47&0.59&\textbf{0.01}&0.66&0.96&0.46&0.08\\
$q_1=4,q_2=3$&0.27&\textbf{0.03}&\textbf{0.00}&0.18&0.76&0.47&0.11&0.16&0.34&0.07&0.59&0.72&\textbf{0.03}&0.82&0.98&0.61&0.43\\
$q_1=8,q_2=6$&0.13&\textbf{0.00}&\textbf{0.00}&0.11&0.62&0.26&0.05&0.07&0.3&\textbf{0.04}&0.51&0.58&\textbf{0.05}&0.80&0.98&0.46&0.2\\
$q_1=12,q_2=9$&0.15&\textbf{0.00}&\textbf{0.00}&\textbf{0.05}&0.57&0.25&\textbf{0.01}&\textbf{0.02}&0.46&\textbf{0.02}&0.42&0.59&\textbf{0.01}&0.80&0.96&0.42&0.13\\
$q_1=16,q_2=12$&0.11&\textbf{0.00}&\textbf{0.00}&0.07&0.49&0.22&\textbf{0.03}&\textbf{0.03}&0.36&\textbf{0.01}&0.58&0.48&\textbf{0.02}&0.68&0.96&0.25&0.09\\
$q_1=4,q_2=2$&0.26&0.08&0.07&0.35&0.47&0.44&0.17&0.31&0.47&0.23&0.69&0.59&0.19&0.80&0.98&0.72&0.42\\
$q_1=8,q_2=4$&0.21&\textbf{0.01}&\textbf{0.01}&0.08&0.57&0.35&\textbf{0.02}&0.09&0.32&0.05&0.50&0.52&\textbf{0.05}&0.78&0.98&0.51&0.29\\
$q_1=12,q_2=6$&0.12&\textbf{0.00}&\textbf{0.00}&0.10&0.62&0.24&0.05&0.05&0.27&\textbf{0.04}&0.49&0.56&\textbf{0.04}&0.80&0.97&0.44&0.19\\
$q_1=16,q_2=8$&0.12&\textbf{0.00}&\textbf{0.00}&\textbf{0.02}&0.56&0.22&\textbf{0.01}&0.08&0.33&\textbf{0.01}&0.35&0.56&\textbf{0.02}&0.75&0.97&0.35&0.11\\
$q_1=8,q_2=2$&0.21&0.07&0.07&0.33&0.42&0.42&0.16&0.28&0.42&0.21&0.68&0.57&0.18&0.78&0.98&0.72&0.42\\
$q_1=12,q_2=3$&0.22&\textbf{0.03}&\textbf{0.00}&0.16&0.75&0.45&0.11&0.10&0.24&0.05&0.55&0.71&\textbf{0.02}&0.80&0.97&0.58&0.42\\
$q_1=16,q_2=4$&0.19&\textbf{0.01}&\textbf{0.01}&0.07&0.56&0.34&\textbf{0.02}&0.07&0.28&\textbf{0.05}&0.48&0.5&\textbf{0.04}&0.76&0.98&0.5&0.28\\
permutation&0.22&\textbf{0.00}&\textbf{0.00}&\textbf{0.04}&0.49&0.18&\textbf{0.00}&0.12&0.46&\textbf{0.04}&0.50&0.39&\textbf{0.01}&0.85&0.94&0.33&0.08\\
bootstrap&0.18&\textbf{0.00}&\textbf{0.00}&\textbf{0.04}&0.50&0.19&\textbf{0.00}&0.11&0.44&\textbf{0.02}&0.54&0.44&\textbf{0.01}&0.88&0.99&0.36&0.09\\
  \bottomrule
\end{tabularx}
\end{table}
\end{landscape}

\begin{landscape}
\begin{table}[!h]
\centering
\scriptsize
\captcont*{Empirical results: p-values, Gini measure, Continued\label{tab_emp_3}}
\begin{tabularx}{1.12\textwidth}{ccccccccccccccccccccccccccccc} \toprule
&\begin{sideways}Magadan Region\end{sideways}&\begin{sideways}Moscow city\end{sideways}&\begin{sideways}Moscow region\end{sideways}&\begin{sideways}Murmansk region\end{sideways}&\begin{sideways}Novgorod region\end{sideways}&\begin{sideways}Novosibirsk region\end{sideways}&\begin{sideways}Omsk region\end{sideways}&\begin{sideways}Orenburg region\end{sideways}&\begin{sideways}Oryol Region\end{sideways}&\begin{sideways}Penza region\end{sideways}&\begin{sideways}Perm region\end{sideways}&\begin{sideways}Pskov region\end{sideways}&\begin{sideways}Rostov region\end{sideways}&\begin{sideways}Ryazan Oblast\end{sideways}&\begin{sideways}Saratov region\end{sideways}&\begin{sideways}Sakhalin Region\end{sideways}&\begin{sideways}Sverdlovsk region\end{sideways}\\\hline
Gini&0.303&0.264&0.298&0.290&0.252&0.305&0.294&0.270&0.257&0.248&0.273&0.259&0.262&0.258&0.258&0.326&0.278\\
$N_2$&720&9000&5952&1296&1176&2760&2088&2160&1248&1656&2736&1248&3816&1488&2640&1080&4152\\
$N_1/N_2$&12.50&1.00&1.51&6.94&7.65&3.26&4.31&4.17&7.21&5.43&3.29&7.21&2.36&6.05&3.41&8.33&2.17\\\hline
$\zeta$&4.14& 3.96&4.70&4.21&4.77&3.50&3.45&4.53&3.11&3.06&4.73&3.97&4.72&3.06&4.24&3.59&3.24\\\hline
asymptotic&\textbf{0.00}&&\textbf{0.00}&\textbf{0.00}&0.06&\textbf{0.00}&\textbf{0.00}&0.27&0.43&\textbf{0.03}&0.07&0.49&0.57&0.47&0.22&\textbf{0.00}&\textbf{0.02}\\
$q_1=q_2=4$&\textbf{0.01}&&\textbf{0.01}&0.06&0.11&\textbf{0.00}&\textbf{0.02}&0.41&0.47&0.19&0.23&0.37&0.64&0.55&0.34&\textbf{0.01}&0.12\\
$q_1=q_2=8$&\textbf{0.00}&&\textbf{0.00}&\textbf{0.01}&\textbf{0.03}&\textbf{0.00}&\textbf{0.00}&0.35&0.42&0.09&0.18&0.43&0.58&0.48&0.23&\textbf{0.00}&\textbf{0.04}\\
$q_1=q_2=12$&\textbf{0.00}&&\textbf{0.00}&\textbf{0.02}&0.07&\textbf{0.00}&\textbf{0.00}&0.32&0.44&0.08&0.11&0.4&0.57&0.44&0.24&\textbf{0.00}&\textbf{0.05}\\
$q_1=q_2=16$&\textbf{0.00}&&\textbf{0.00}&\textbf{0.01}&\textbf{0.02}&\textbf{0.00}&\textbf{0.00}&0.31&0.39&\textbf{0.03}&0.08&0.35&0.52&0.37&0.17&\textbf{0.00}&0.06\\
$q_1=4,q_2=3$&\textbf{0.03}&&\textbf{0.04}&0.12&0.16&\textbf{0.02}&\textbf{0.02}&0.58&0.52&0.35&0.31&0.57&0.69&0.31&0.47&\textbf{0.03}&0.23\\
$q_1=8,q_2=6$&\textbf{0.00}&&\textbf{0.00}&\textbf{0.03}&0.11&\textbf{0.00}&\textbf{0.00}&0.38&0.49&0.14&0.17&0.40&0.59&0.49&0.31&\textbf{0.00}&0.07\\
$q_1=12,q_2=9$&\textbf{0.00}&&\textbf{0.00}&\textbf{0.04}&0.07&\textbf{0.00}&\textbf{0.00}&0.28&0.44&0.10&0.11&0.44&0.62&0.45&0.23&\textbf{0.00}&\textbf{0.04}\\
$q_1=16,q_2=12$&\textbf{0.00}&&\textbf{0.00}&\textbf{0.02}&0.07&\textbf{0.00}&\textbf{0.00}&0.32&0.43&0.08&0.11&0.39&0.56&0.43&0.23&\textbf{0.00}&0.05\\
$q_1=4,q_2=2$&0.09&&0.07&0.21&0.23&0.08&0.15&0.61&0.5&0.34&0.48&0.42&0.79&0.64&0.49&\textbf{0.04}&0.30\\
$q_1=8,q_2=4$&\textbf{0.01}&&\textbf{0.01}&0.06&0.09&\textbf{0.00}&\textbf{0.02}&0.39&0.46&0.18&0.21&0.31&0.61&0.54&0.31&\textbf{0.01}&0.11\\
$q_1=12,q_2=6$&\textbf{0.00}&&\textbf{0.00}&\textbf{0.02}&0.11&\textbf{0.00}&\textbf{0.00}&0.37&0.49&0.14&0.16&0.38&0.58&0.48&0.3&\textbf{0.00}&0.06\\
$q_1=16,q_2=8$&\textbf{0.00}&&\textbf{0.00}&\textbf{0.01}&\textbf{0.02}&\textbf{0.00}&\textbf{0.00}&0.33&0.41&0.09&0.16&0.42&0.55&0.47&0.2&\textbf{0.00}&\textbf{0.03}\\
$q_1=8,q_2=2$&0.08&&0.06&0.20&0.20&0.08&0.14&0.60&0.48&0.34&0.47&0.35&0.78&0.63&0.46&\textbf{0.03}&0.28\\
$q_1=12,q_2=3$&\textbf{0.02}&&\textbf{0.03}&0.11&0.13&\textbf{0.02}&\textbf{0.01}&0.57&0.50&0.35&0.29&0.55&0.66&0.25&0.45&\textbf{0.03}&0.21\\
$q_1=16,q_2=4$&\textbf{0.01}&&\textbf{0.00}&0.06&0.08&\textbf{0.00}&\textbf{0.02}&0.38&0.44&0.17&0.2&0.26&0.58&0.53&0.29&\textbf{0.01}&0.10\\
permutation&\textbf{0.00}&&\textbf{0.00}&\textbf{0.00}&0.06&\textbf{0.00}&\textbf{0.00}&0.25&0.45&\textbf{0.02}&0.07&0.5&0.58&0.52&0.21&\textbf{0.00}&\textbf{0.03}\\
bootstrap&\textbf{0.00}&&\textbf{0.00}&\textbf{0.00}&0.08&\textbf{0.00}&\textbf{0.00}&0.26&0.41&\textbf{0.03}&0.07&0.52&0.6&0.46&0.23&\textbf{0.00}&\textbf{0.02}\\
  \bottomrule
\end{tabularx}
\end{table}
\end{landscape}

\begin{landscape}
\begin{table}[!h]
\centering
\scriptsize
\captcont*{Empirical results: p-values, Gini measure, Continued\label{tab_emp_4}}
\begin{tabularx}{1.12\textwidth}{ccccccccccccccccccccccccccccc} \toprule
&\begin{sideways}Smolensk region\end{sideways}&\begin{sideways}Sevastopol city\end{sideways}&\begin{sideways}Tambov Region\end{sideways}&\begin{sideways}Tomsk region\end{sideways}&\begin{sideways}Tula region\end{sideways}&\begin{sideways}Tyumen region\end{sideways}&\begin{sideways}Khanty-Mansi Autonomous Okrug - Yugra\end{sideways}&\begin{sideways}Ulyanovsk region\end{sideways}&\begin{sideways}Yamalo-Nenets Autonomous District\end{sideways}&\begin{sideways}Chelyabinsk region\end{sideways}&\begin{sideways}Zabaykalsky Krai\end{sideways}&\begin{sideways}Chukotka Autonomous District\end{sideways}&\begin{sideways}Yaroslavskaya oblast\end{sideways}&\begin{sideways}Republic of Adygea\end{sideways}&\begin{sideways}Republic of Bashkortostan\end{sideways}&\begin{sideways}The Republic of Buryatia\end{sideways}&\begin{sideways}The Republic of Dagestan\end{sideways}\\\hline
Gini&0.251&0.240&0.236&0.315&0.248&0.324&0.293&0.245&0.335&0.283&0.317&0.328&0.260&0.267&0.292&0.327&0.318\\
$N_2$&1440&792&1488&1416&1848&1560&1656&1560&1008&3408&1368&528&1632&936&3624&1272&1968\\
$N_1/N_2$&6.25&11.36&6.05&6.36&4.87&5.77&5.43&5.77&8.93&2.64&6.58&17.05&5.51&9.62&2.48&7.08&4.57\\\hline
$\zeta$&4.41&5.78&4.32&3.13&5.40&2.71&4.11&4.36&5.46&4.00&5.17&4.64&4.19&5.46&4.48&3.53&4.26\\\hline
asymptotic&\textbf{0.05}&\textbf{0.00}&\textbf{0.00}&\textbf{0.00}&\textbf{0.00}&\textbf{0.00}&\textbf{0.00}&\textbf{0.00}&\textbf{0.00}&\textbf{0.00}&\textbf{0.00}&\textbf{0.00}&0.46&0.7&\textbf{0.00}&\textbf{0.00}&\textbf{0.00}\\
$q_1=q_2=4$&0.12&\textbf{0.03}&\textbf{0.02}&\textbf{0.01}&\textbf{0.03}&\textbf{0.01}&\textbf{0.01}&0.08&\textbf{0.00}&\textbf{0.02}&\textbf{0.02}&\textbf{0.00}&0.42&0.83&\textbf{0.01}&\textbf{0.01}&\textbf{0.03}\\
$q_1=q_2=8$&0.08&\textbf{0.01}&\textbf{0.00}&\textbf{0.00}&\textbf{0.01}&\textbf{0.00}&\textbf{0.00}&\textbf{0.03}&\textbf{0.00}&\textbf{0.02}&\textbf{0.00}&\textbf{0.00}&0.26&0.81&\textbf{0.00}&\textbf{0.00}&\textbf{0.00}\\
$q_1=q_2=12$&\textbf{0.03}&\textbf{0.00}&\textbf{0.00}&\textbf{0.00}&\textbf{0.01}&\textbf{0.00}&\textbf{0.00}&\textbf{0.01}&\textbf{0.00}&\textbf{0.00}&\textbf{0.00}&\textbf{0.00}&0.28&0.97&\textbf{0.00}&\textbf{0.00}&\textbf{0.00}\\
$q_1=q_2=16$&\textbf{0.04}&\textbf{0.01}&\textbf{0.00}&\textbf{0.00}&\textbf{0.00}&\textbf{0.00}&\textbf{0.00}&\textbf{0.00}&\textbf{0.00}&\textbf{0.00}&\textbf{0.00}&\textbf{0.00}&0.30&0.78&\textbf{0.00}&\textbf{0.00}&\textbf{0.00}\\
$q_1=4,q_2=3$&0.24&0.05&\textbf{0.02}&\textbf{0.01}&0.08&\textbf{0.01}&\textbf{0.02}&0.12&\textbf{0.01}&\textbf{0.04}&0.07&\textbf{0.02}&0.33&0.57&0.06&\textbf{0.01}&0.06\\
$q_1=8,q_2=6$&0.08&\textbf{0.02}&\textbf{0.00}&\textbf{0.00}&\textbf{0.01}&\textbf{0.00}&\textbf{0.00}&0.06&\textbf{0.00}&\textbf{0.01}&\textbf{0.01}&\textbf{0.00}&0.3&0.8&\textbf{0.00}&\textbf{0.00}&\textbf{0.00}\\
$q_1=12,q_2=9$&0.11&\textbf{0.00}&\textbf{0.00}&\textbf{0.00}&\textbf{0.01}&\textbf{0.00}&\textbf{0.00}&\textbf{0.02}&\textbf{0.00}&\textbf{0.00}&\textbf{0.00}&\textbf{0.00}&0.33&0.8&\textbf{0.00}&\textbf{0.00}&\textbf{0.00}\\
$q_1=16,q_2=12$&\textbf{0.03}&\textbf{0.00}&\textbf{0.00}&\textbf{0.00}&\textbf{0.01}&\textbf{0.00}&\textbf{0.00}&\textbf{0.01}&\textbf{0.00}&\textbf{0.00}&\textbf{0.00}&\textbf{0.00}&0.27&0.98&\textbf{0.00}&\textbf{0.00}&\textbf{0.00}\\
$q_1=4,q_2=2$&0.29&0.1&0.12&0.08&0.15&0.07&0.11&0.23&0.1&0.13&0.08&\textbf{0.04}&0.45&0.56&0.12&0.07&0.19\\
$q_1=8,q_2=4$&0.11&\textbf{0.02}&\textbf{0.02}&\textbf{0.01}&\textbf{0.02}&\textbf{0.01}&\textbf{0.00}&0.07&\textbf{0.00}&\textbf{0.01}&\textbf{0.02}&\textbf{0.00}&0.38&0.82&\textbf{0.01}&\textbf{0.01}&\textbf{0.03}\\
$q_1=12,q_2=6$&0.07&\textbf{0.02}&\textbf{0.00}&\textbf{0.00}&\textbf{0.01}&\textbf{0.00}&\textbf{0.00}&0.05&\textbf{0.00}&\textbf{0.01}&\textbf{0.01}&\textbf{0.00}&0.27&0.79&\textbf{0.00}&\textbf{0.00}&\textbf{0.00}\\
$q_1=16,q_2=8$&0.07&\textbf{0.01}&\textbf{0.00}&\textbf{0.00}&\textbf{0.01}&\textbf{0.00}&\textbf{0.00}&\textbf{0.03}&\textbf{0.00}&\textbf{0.01}&\textbf{0.00}&\textbf{0.00}&0.21&0.81&\textbf{0.00}&\textbf{0.00}&\textbf{0.00}\\
$q_1=8,q_2=2$&0.28&0.08&0.11&0.08&0.13&0.07&0.1&0.22&0.1&0.12&0.08&\textbf{0.04}&0.39&0.5&0.11&0.07&0.19\\
$q_1=12,q_2=3$&0.22&\textbf{0.04}&\textbf{0.02}&\textbf{0.01}&0.06&\textbf{0.01}&\textbf{0.01}&0.11&\textbf{0.00}&\textbf{0.03}&0.07&\textbf{0.02}&0.20&0.46&0.05&\textbf{0.01}&0.06\\
$q_1=16,q_2=4$&0.10&\textbf{0.02}&\textbf{0.02}&\textbf{0.01}&\textbf{0.02}&\textbf{0.01}&\textbf{0.00}&0.06&\textbf{0.00}&\textbf{0.01}&\textbf{0.02}&\textbf{0.00}&0.35&0.82&\textbf{0.01}&\textbf{0.01}&\textbf{0.03}\\
permutation&0.08&\textbf{0.00}&\textbf{0.00}&\textbf{0.00}&\textbf{0.01}&\textbf{0.00}&\textbf{0.00}&\textbf{0.00}&\textbf{0.00}&\textbf{0.00}&\textbf{0.00}&\textbf{0.00}&0.51&0.70&\textbf{0.00}&\textbf{0.00}&\textbf{0.00}\\
bootstrap&0.07&\textbf{0.01}&\textbf{0.00}&\textbf{0.00}&\textbf{0.00}&\textbf{0.00}&\textbf{0.00}&\textbf{0.01}&\textbf{0.00}&\textbf{0.00}&\textbf{0.00}&\textbf{0.00}&0.47&0.63&\textbf{0.00}&\textbf{0.00}&\textbf{0.00}\\
  \bottomrule
\end{tabularx}
\end{table}
\end{landscape}

\begin{landscape}
\begin{table}[!h]
\centering
\scriptsize
\captcont*{Empirical results: p-values, Gini measure, Continued\label{tab_emp_5}}
\begin{tabularx}{1.12\textwidth}{ccccccccccccccccccccccccccccc} \toprule
&\begin{sideways}Kabardino-Balkar Republic\end{sideways}&\begin{sideways}Altai Republic\end{sideways}&\begin{sideways}Republic of Kalmykia\end{sideways}&\begin{sideways}Republic of Karelia\end{sideways}&\begin{sideways}Komi Republic\end{sideways}&\begin{sideways}Mari El Republic\end{sideways}&\begin{sideways}The Republic of Mordovia\end{sideways}&\begin{sideways}Republic of North Ossetia-Alania\end{sideways}&\begin{sideways}Karachay-Cherkess Republic\end{sideways}&\begin{sideways}Republic of Tatarstan\end{sideways}&\begin{sideways}Tyva Republic\end{sideways}&\begin{sideways}Udmurtia\end{sideways}&\begin{sideways}The Republic of Khakassia\end{sideways}&\begin{sideways}Chechen Republic\end{sideways}&\begin{sideways}Chuvash Republic - Chuvashia\end{sideways}&\begin{sideways}The Republic of Sakha (Yakutia)\end{sideways}&\begin{sideways}Jewish Autonomous Region\end{sideways}\\\hline
Gini&0.298&0.340&0.326&0.268&0.298&0.286&0.251&0.277&0.334&0.248&0.423&0.251&0.293&0.327&0.267&0.341&0.334\\
$N_2$&960&600&720&1200&1392&1104&1248&912&720&3408&744&1704&1032&1032&1488&1320&600\\
$N_1/N_2$&9.38&15.00&12.50&7.50&6.47&8.15&7.21&9.87&12.50&2.64&12.10&5.28&8.72&8.72&6.05&6.82&15.00\\\hline
$\zeta$&4.28&3.35&3.47&3.78&4.14&2.29&3.12&3.34&2.08&4.49&3.04&4.50&5.14&3.30&5.27&4.34&4.27\\\hline
asymptotic&\textbf{0.00}&\textbf{0.00}&\textbf{0.00}&0.64&\textbf{0.00}&0.06&0.14&0.18&\textbf{0.00}&\textbf{0.00}&\textbf{0.00}&\textbf{0.03}&\textbf{0.00}&\textbf{0.00}&0.62&\textbf{0.00}&\textbf{0.00}\\
$q_1=q_2=4$&\textbf{0.04}&\textbf{0.00}&\textbf{0.01}&0.71&\textbf{0.03}&0.14&0.28&0.2&\textbf{0.03}&\textbf{0.03}&\textbf{0.00}&0.11&\textbf{0.03}&\textbf{0.02}&0.63&\textbf{0.00}&\textbf{0.05}\\
$q_1=q_2=8$&\textbf{0.01}&\textbf{0.00}&\textbf{0.00}&0.65&\textbf{0.00}&0.15&0.2&0.21&\textbf{0.01}&\textbf{0.01}&\textbf{0.00}&\textbf{0.02}&\textbf{0.01}&\textbf{0.00}&0.63&\textbf{0.00}&\textbf{0.01}\\
$q_1=q_2=12$&\textbf{0.00}&\textbf{0.00}&\textbf{0.00}&0.58&\textbf{0.00}&0.12&0.11&0.27&\textbf{0.01}&\textbf{0.00}&\textbf{0.00}&\textbf{0.02}&\textbf{0.00}&\textbf{0.00}&0.64&\textbf{0.00}&\textbf{0.00}\\
$q_1=q_2=16$&\textbf{0.00}&\textbf{0.00}&\textbf{0.00}&0.63&\textbf{0.00}&0.16&0.14&0.23&\textbf{0.01}&\textbf{0.00}&\textbf{0.00}&0.08&\textbf{0.00}&\textbf{0.00}&0.68&\textbf{0.00}&\textbf{0.00}\\
$q_1=4,q_2=3$&\textbf{0.02}&\textbf{0.00}&\textbf{0.03}&0.68&0.06&0.13&0.22&0.43&\textbf{0.01}&0.13&\textbf{0.01}&0.08&0.09&0.05&0.55&\textbf{0.00}&0.10\\
$q_1=8,q_2=6$&\textbf{0.00}&\textbf{0.00}&\textbf{0.00}&0.59&\textbf{0.00}&\textbf{0.03}&0.2&0.32&\textbf{0.01}&\textbf{0.03}&\textbf{0.00}&0.08&\textbf{0.01}&\textbf{0.01}&0.6&\textbf{0.00}&\textbf{0.01}\\
$q_1=12,q_2=9$&\textbf{0.00}&\textbf{0.00}&\textbf{0.00}&0.62&\textbf{0.00}&0.14&0.12&0.33&\textbf{0.00}&\textbf{0.01}&\textbf{0.00}&0.07&\textbf{0.01}&\textbf{0.00}&0.8&\textbf{0.00}&\textbf{0.01}\\
$q_1=16,q_2=12$&\textbf{0.00}&\textbf{0.00}&\textbf{0.00}&0.59&\textbf{0.00}&0.12&0.11&0.27&\textbf{0.01}&\textbf{0.00}&\textbf{0.00}&\textbf{0.02}&\textbf{0.00}&\textbf{0.00}&0.65&\textbf{0.00}&\textbf{0.00}\\
$q_1=4,q_2=2$&0.16&\textbf{0.03}&0.14&0.48&0.16&0.16&0.18&0.21&\textbf{0.05}&0.18&0.13&0.28&0.12&\textbf{0.05}&0.54&0.06&0.14\\
$q_1=8,q_2=4$&\textbf{0.04}&\textbf{0.00}&\textbf{0.01}&0.71&\textbf{0.03}&0.13&0.27&0.18&\textbf{0.03}&\textbf{0.02}&\textbf{0.00}&0.09&\textbf{0.03}&\textbf{0.02}&0.62&\textbf{0.00}&\textbf{0.05}\\
$q_1=12,q_2=6$&\textbf{0.00}&\textbf{0.00}&\textbf{0.00}&0.58&\textbf{0.00}&\textbf{0.03}&0.19&0.31&\textbf{0.01}&\textbf{0.02}&\textbf{0.00}&0.07&\textbf{0.01}&\textbf{0.01}&0.57&\textbf{0.00}&\textbf{0.01}\\
$q_1=16,q_2=8$&\textbf{0.01}&\textbf{0.00}&\textbf{0.00}&0.65&\textbf{0.00}&0.15&0.19&0.21&\textbf{0.01}&\textbf{0.01}&\textbf{0.00}&\textbf{0.01}&\textbf{0.00}&\textbf{0.00}&0.61&\textbf{0.00}&\textbf{0.01}\\
$q_1=8,q_2=2$&0.15&\textbf{0.03}&0.13&0.41&0.15&0.15&0.14&0.19&\textbf{0.04}&0.16&0.13&0.26&0.12&\textbf{0.04}&0.48&0.06&0.14\\
$q_1=12,q_2=3$&\textbf{0.01}&\textbf{0.00}&\textbf{0.03}&0.66&0.06&0.12&0.20&0.42&\textbf{0.01}&0.12&\textbf{0.01}&\textbf{0.05}&0.09&0.05&0.45&\textbf{0.00}&0.10\\
$q_1=16,q_2=4$&\textbf{0.03}&\textbf{0.00}&\textbf{0.01}&0.71&\textbf{0.03}&0.13&0.26&0.17&\textbf{0.03}&\textbf{0.02}&\textbf{0.00}&0.08&\textbf{0.03}&\textbf{0.02}&0.61&\textbf{0.00}&\textbf{0.05}\\
permutation&\textbf{0.00}&\textbf{0.00}&\textbf{0.00}&0.59&\textbf{0.00}&\textbf{0.04}&0.18&0.18&\textbf{0.00}&\textbf{0.00}&\textbf{0.00}&\textbf{0.04}&\textbf{0.00}&\textbf{0.00}&0.6&\textbf{0.00}&\textbf{0.00}\\
bootstrap&\textbf{0.00}&\textbf{0.00}&\textbf{0.00}&0.63&\textbf{0.00}&\textbf{0.05}&0.18&0.17&\textbf{0.00}&\textbf{0.00}&\textbf{0.00}&0.05&\textbf{0.00}&\textbf{0.00}&0.64&\textbf{0.00}&\textbf{0.00}\\
  \bottomrule
\end{tabularx}
\end{table}
\end{landscape}

\newpage \setcounter{section}{1}
\appendix \label{AppB} {\centerline{\textbf{\large{Appendix B: Inequality measures and their sample analogues }}}} 
 \renewcommand\theequation{B.\arabic{equation}}
\par \bigskip
In this section, we review the definitions of the widely used Gini and Theil inequality measures, sample analogues of the measures and their asymptotic properties (see, among others, \cite{CF, DF}, Section 13.F, 17.C in \cite{MO}, and references therein).

Let $I$ be an (absolutely continuous) nonnegative r.v. (e.g., income or wealth level) with the finite first moment $\mu_I=E[I]<\infty$ and the cdf $F_I(x)$ representing income or wealth distribution in a population, and let $I_1, I_2, ..., I_N$ denote a sample of observations on the r.v. $I.$

As usual, we denote by $\overline{I}_N=N^{-1}\sum_{i=1}^N I_i$ and $s_N^2=(N-1)^{-1}\sum_{i=1}^N (I_i-\overline{I})^2$ the sample mean and sample variance of the observations $I_i.$

Below, we provide the definitions of Theil and Gini inequality measures (denoted by $\mathcal{L}_{Theil}^I$ and $\mathcal{L}_{Gini}^I$ for the population considered) and discuss the standard 
results on their asymptotic normality.

\textbf{Theil index}  The population Theil index is defined by 
$$\mathcal{L}_{Theil}^I=\frac{E[I \log I]}{\mu_I}-\log(\mu_I).$$

The Theil index is the limiting case of the Generalized Entropy measures. 
Its sample analogue - sample Theil index - is given by

$$\hat{\mathcal{L}}_{Theil, N}^I=\frac{\frac{1}{N}\sum_{i=1}^N I_i \log(I_i)}{\overline{I}_N}-
\log(\overline{I}_N).$$

Under i.i.d. observations $I_1, I_2, ..., I_N,$ the Theil index is asymptotically normal if $E[I^2]<\infty,$ $E[I^2\log I]<\infty$ and $E[I^2 \log^2(I)]<\infty.$ It is easy to see that these conditions are satisfied in the case of r.v.'s with power law distributions (\ref{power1}) (e.g., Singh-Maddala distributions $SM(a, b, c)$ in (\ref{SM}) with $\zeta=ac$) if the tail index $\zeta$ is greater than 2: $\zeta>2.$ 

Under the above conditions, one has 

$$\sqrt{N}(\hat{\mathcal{L}}_{Theil, N}^I-\mathcal{L}_{Theil}^I)\rightarrow_w N(0, v_{Theil, I}^2),$$

where $$v_{Theil, I}^2=\frac{E[I^2 \log^2 I]}{\mu_I^2}+\frac{E[I^2]} {\mu_I^2} \Big(\frac{E[I \log I]}{\mu_I}+1\Big)^2-\frac{2E[I^2 \log I]}{\mu_I^2}\Big(\frac{E[I \log I]}{\mu_I}+1\Big)-1$$

\noindent (see, among others, \cite{MZ}, \cite{Cow1, Cow2}, \cite{CF} and \cite{Afr} for the review of the results on asymptotic normality and the formulas for the liming and sampling variance of different estimators of inequality measures).

\textbf{Gini coefficient} The population Gini coefficient is defined by $$\mathcal{L}_{Gini}^I=0.5 \frac{E|I'-I''|}{\mu_I},$$ where $I'$ and $I''$ are independent copies of the r.v. $I.$

The most commonly used (nonparametric) estimator of the Gini coefficient $\mathcal{L}_{Gini}^I$ is given by its sample analogue (the sample Gini coefficient)
\begin{eqnarray} \label{sampleG} \nonumber \hat{\mathcal{L}}_{Gini, N}^I=\frac{\sum_{1\le i<j\le N} |I_i-I_j|}{(N-1)\sum_{i=1}^N I_i}=U_N/\overline{I}_N, \end{eqnarray}
where $U_N$ is the $U-$statistic $U_N=\frac{2}{N(N-1)} \sum_{1\le i<j\le N}  |I_i-I_j|$ (we refer to, among others, \cite{Hoeffding}, Ch. 5 in \cite{Serfling} and Ch. 4 in \cite{KB} for the asymptotic theory for general $U-$statistics).

From the results in the above references, it follows that asymptotic normality for the $U-$statistic $U_N$ and the sample Gini coefficient holds if $I_1, I_2, ..., I_N$ are i.i.d. observations with finite second moment $E[I^2]<\infty.$ This holds under power-law distributions (\ref{power1}) (e.g., for Singh-Maddala distributions $SM(a, b, c)$ in (\ref{SM}) with $\zeta=ac$) if the tail index $\zeta$ is greater than 2: $\zeta>2.$ More precisely, under the above conditions (see \cite{Hoeffding})
$$\sqrt{N}(\hat{\mathcal{L}}_{Gini, N}^I-\mathcal{L}_{Gini}^I)\rightarrow_w N(0, v_{Gini, I}^2),$$
where $v_{Gini, I}^2=(\mathcal{L}_{Gini}^I)^2 \sigma_I^2-2 \mathcal{L}_{Gini}^I E\{I'|I'-I''|\}/\mu_I^2+E(E_{I'}\{|I'-I''|\})/\mu_I^2,$ and $E_{I'}(\cdot)=E_{I'}(\cdot)=E\{\cdot|I'\}$ denotes the expectation conditional on $I'.$

Naturally, the asymptotic normality of the sample Theil and Gini  coefficients is lost under infinite second moments and variances: $E[I^2]=\infty.$ For instance, from the results in \cite{Taleb} it follows that under i.i.d. observations $I_1, I_2, ..., I_N$ that follow a power-law distribution (\ref{power1}) with the tail index $\zeta\in (1, 2)$ (e.g., the Singh-Maddala distribution $SM(a, b, c)$ in (\ref{SM}) with $1<\zeta=ac<2$) and have finite first and infinite second moments, the sample Gini coefficient $\hat{\mathcal{L}}_{Gini, N}$ has an asymptotic right-skewed stable distribution with the index of stability $\zeta.$ Using the standard generalized CLT and the delta-method, it is also not difficult to show that in the case of distributions exhibiting (double) power law behavior in both the lower and the upper (with the tail index $\zeta$), similar to Singh-Maddala distributions $SM(a, b, c)$ with $\zeta=ac,$ the sample Theil index $\hat{\mathcal{L}}_{Theil, N}$ weakly converges to a function of stable r.v.'s with indices of stability that depend on $\zeta.$ The rate of convergence in the above asymptotic results is slower than $\sqrt{N}$ and depends on $\zeta.$ The fact that the tail index $\zeta$ is unknown in practice makes the results useless for (direct) asymptotic inference.\footnote{The situation is somewhat similar to the properties of autocorrelation functions of GARCH-type processes and their squares, where asymptotic normality is lost under tail indices smaller than 4 and infinite fourth moments, as is typically the case for financial returns and foreign exchange rates in real-world markets (see  \cite{DM}, \cite{MS} and also \cite{IPS} for asymptotically valid robust $t-$statistic approaches to inference on measures of market (non-)efficiency and volatility clustering based on powers of absolute values of GARCH-type processes, e.g., financial returns).}
\end{document}